\documentclass[12pt]{article}
\pdfoutput=1
\usepackage{jheppub}
\usepackage{epsfig}
\usepackage{amsmath}
\usepackage{amssymb}
\usepackage{amsfonts}
\usepackage{amsxtra}
\usepackage{amsthm}
\usepackage{mathrsfs}
\usepackage{mathdots}
\usepackage{makeidx}
\usepackage{graphics}
\usepackage{dsfont}
\usepackage{mathtools}
\usepackage{graphicx}
\usepackage{placeins}
\usepackage{bm}
\usepackage[capitalize]{cleveref}
\usepackage{empheq}
\usepackage{colortbl}
\usepackage{xcolor}
\usepackage{enumerate}
\usepackage{titlesec}
\usepackage{longtable}
\usepackage{hhline}
\usepackage{float}
\usepackage{color}
\usepackage{tikz}
\usepackage{xfrac}
\usepackage{footnote}
\usepackage{rotating}
\usepackage{lscape}
\usepackage{makecell}
\usepackage{environ}
\usepackage{tabularx}
\usepackage{subfiles}
\usepackage[export]{adjustbox}
\usepackage{ytableau, youngtab}
\usepackage{tikz-3dplot}
\usepackage{slashed}
\usepackage{pifont}
\usepackage{multirow}
\usepackage{mdframed}
\usepackage{bbm}
\usepackage[normalem]{ulem}
\usepackage{arydshln}
\usepackage[inline]{enumitem} 
\usepackage{fancybox}
\usepackage[numbers,compress]{natbib}
\usepackage{subfiles}
\usepackage{crossreftools}
\usepackage{pdflscape}
\usepackage{hhline}
\usepackage{orcidlink}

\usetikzlibrary{positioning,trees,decorations.pathmorphing,decorations.markings,decorations.pathreplacing,calc,shapes,patterns,arrows,chains,arrows.meta,fit,fadings,decorations.markings,graphs,graphs.standard,quotes,plotmarks,calligraphy}

\definecolor{prhigh}{HTML}{c2882b}
\definecolor{prcolor}{HTML}{174840}
\definecolor{seccolor}{HTML}{22624b}
\definecolor{tercolor}{HTML}{a24400}

\DeclareMathOperator{\U}{U}
\DeclareMathOperator{\SU}{SU}

\newcommand{\cN}{\mathcal{N}}

\newcommand{\coma}{\, , \quad}
\newcommand{\fstop}{\, .}

\newcommand{\aand}{\quad\text{and}\quad}

\renewcommand{\epsilon}{\varepsilon}

\theoremstyle{plain}

\theoremstyle{definition}

\let\oldFootnote\footnote
\newcommand\nextToken\relax

\renewcommand\footnote[1]{%
	\oldFootnote{#1}\futurelet\nextToken\isFootnote}

\newcommand\isFootnote{%
	\ifx\footnote\nextToken\textsuperscript{,}\fi}

\tikzset{cross/.style={cross out, draw=black, fill=none, minimum size=2*(#1-\pgflinewidth), inner sep=0pt, outer sep=0pt}, cross/.default={2pt}}

\newcommand{\Lpagenumber}{\ifdim\textwidth=\linewidth\else\bgroup
	\dimendef\margin=0 
	\ifodd\value{page}\margin=\oddsidemargin
	\else\margin=\evensidemargin
	\fi
	\raisebox{\dimexpr -3\topmargin-\headheight-\headsep-0.5\linewidth}[0pt][0pt]{%
		\rlap{\hspace{\dimexpr \margin+\textheight+2\footskip}%
			\llap{\rotatebox{90}{\hfill-- \thepage\ --\hfill}}}}%
	\egroup\fi}
\AddToHook{shipout/background}{\Lpagenumber}%

\usetikzlibrary{positioning}
\usetikzlibrary{chains}
\usetikzlibrary{arrows, arrows.meta ,fit,decorations.pathreplacing}
\tikzstyle{every picture}+=[remember picture]
\tikzstyle{na} = [baseline]
\tikzstyle{ligne}=[draw, thick]

\usetikzlibrary{arrows, decorations.markings, calc, fadings, decorations.pathreplacing, patterns, decorations.pathmorphing, positioning}
\tikzset{>={Latex[width=1.5mm,length=1.5mm]}}
\tikzset{bd/.style={circle, draw=black, inner sep=0pt, fill=black, minimum size=1.2mm}}
\tikzset{bld/.style={circle, draw=blue, inner sep=0pt, fill=blue, minimum size=1.2mm}}
\tikzset{wd/.style={circle, draw=black, inner sep=0pt, fill=white, minimum size=1.2mm}}
\tikzset{rd/.style={circle, draw=red, inner sep=0pt, fill=red, minimum size=.9mm}}
\tikzset{wrd/.style={circle, draw=red, inner sep=0pt, fill=white, minimum size=.9mm}}
\usetikzlibrary{graphs,graphs.standard,quotes}

\def\node#1#2{\overset{#1}{\underset{#2}{{\color{gray} \bullet}}}}

\def\node#1#2{\overset{#1}{\underset{#2}{\circ}}}

\tikzstyle{every picture}+=[remember picture]
\tikzstyle{na} = [baseline=-.5ex]

\newcommand{\ie}{i.e. }

\numberwithin{equation}{section}
\newcommand{\bes}[1]{\begin{equation} \begin{split} #1\end{split} \end{equation}}

\newcommand{\be}{\begin{equation}} \newcommand{\ee}{\end{equation}}
\newcommand{\bea}{\begin{equation} \begin{aligned}} \newcommand{\eea}{\end{aligned} \end{equation}}

\def\tilde{\widetilde}

\def\bar{\overline}

\def\rt2{\sqrt{2}}

\def\Tr{\mathop{\rm Tr}}

\def\1{{\ds 1}}

\newcommand{\CS}{\mathcal{S}}

\newcommand{\fn}{\mathfrak{n}}

\def\SU{\mathrm{SU}}

\def\fp{\mathfrak{p}}

\def\fh{\mathfrak{h}}

\def\repa{\raise4pt\hbox{$\square$}\mkern-14mu\raise-4pt\hbox{$\square$}}
\def\repab{\overline{\raise4pt\hbox{$\square$}\mkern-14mu\raise-4pt\hbox{$\square$}\mkern-1mu}}

\def\smileface{\ensuremath{\hbox{\large$\bigcirc$}\mkern-15mu\raise-1pt\hbox{\scriptsize$\smallsmile$}%
		\mkern-10mu\raise4pt\hbox{..}\mkern4mu}}
\def\frownface{\ensuremath{\hbox{\large$\bigcirc$}\mkern-15mu\raise-1pt\hbox{\scriptsize$\smallfrown$}%
		\mkern-10mu\raise4pt\hbox{..}\mkern4mu}}

\newcommand{\ba}{\begin{array}}
	
	\newcommand{\bi}{\begin{itemize}}
		\newcommand{\ei}{\end{itemize}}
	
	\def\bea#1\eea{\allowdisplaybreaks \begin{align}#1\end{align}}
	\newcommand{\ben}{\begin{enumerate}}
		\newcommand{\een}{\end{enumerate}}
	\newcommand{\bean}{\begin{eqnarray*}}
		\newcommand{\eean}{\end{eqnarray*}}

	\newcommand{\BC}{\mathbb{C}}

	\newcommand{\BZ}{\mathbb{Z}}

	\definecolor{light-gray}{gray}{0.5}
	\definecolor{new-green}{rgb}{0,0.7,0.3}

	\def\skipper{\hskip.5em\relax}
	\def\Esix#1#2#3#4#5#6{%
		{\text{\small$ \left[ \begin{array}{c@{\skipper}c@{\skipper}c@{\skipper}c@{\skipper}c@{\skipper}c}
					&&#2&& \\
					#1 &  #3 &  #4 & #5 & #6
					\}$ \right]$}}}
		\def\Eseven#1#2#3#4#5#6#7{%
			{\text{\small$ \left[ \begin{array}{c@{\skipper}c@{\skipper}c@{\skipper}c@{\skipper}c@{\skipper}c}
						&&#2&&& \\
						#1 &  #3 &  #4 & #5 & #6&#7
						\}$ \right]$}}}
			\def\Eeight#1#2#3#4#5#6#7#8{%
				{\text{\small$\begin{array}{c@{\skipper}c@{\skipper}c@{\skipper}c@{\skipper}c@{\skipper}c@{\skipper}c@{\skipper}c}
							&&#2&&&& \\
							#1 &  #3 & #4  & #5 & #6 & #7 &  #8
							\}$$}}}
				\def\atwoEsix#1#2#3#4#5#6#7#8{%
					{\text{\small$\left[#1,#2~;~\begin{array}{c@{\skipper}c@{\skipper}c@{\skipper}c@{\skipper}c@{\skipper}c}
								&&#4&& \\
								#3 &  #5 &  #6 & #7 & #8
								\}$\right]$}}}
					\def\aoneEseven#1#2#3#4#5#6#7#8{%
						{\text{\small$\left[#1~;~\begin{array}{c@{\skipper}c@{\skipper}c@{\skipper}c@{\skipper}c@{\skipper}c}
									&&#3&&& \\
									#2 &  #4 &  #5 & #6 & #7 & #8
									\}$\right]$}}}

						\def\aup#1 {\overset{#1}{\uparrow} \, \overset{\tilde{#1}}{\downarrow}}

						\tikzset{snake it/.style={decorate, decoration={snake, amplitude=.4mm, segment length=2mm,
									post length=0mm,pre length=0mm}}}
						
						\newcommand{\GCD}{\mathrm{GCD}}

						\def\u{\mathfrak{u}}
						\DeclareMathAlphabet{\mymathds}{U}{BOONDOX-ds}{m}{n}
						\tikzstyle{double_border} = [draw, double, double distance=1pt]

						\makeatletter
						\newsavebox{\measure@tikzpicture}
						\NewEnviron{scaletikzpicturetowidth}[1]{%
							\def\tikz@width{#1}%
							\begin{lrbox}{\measure@tikzpicture}%
								\BODY
							\end{lrbox}%
							\pgfmathparse{#1/\wd\measure@tikzpicture}%
							\BODY
						}
						\makeatother
						
						\makeatletter
						\def\squarecorner#1{
							\pgf@x=\the\wd\pgfnodeparttextbox%
							\pgfmathsetlength\pgf@xc{\pgfkeysvalueof{/pgf/inner xsep}}%
							\advance\pgf@x by 2\pgf@xc%
							\pgfmathsetlength\pgf@xb{\pgfkeysvalueof{/pgf/minimum width}}%
							\ifdim\pgf@x<\pgf@xb%
							\pgf@x=\pgf@xb%
							\fi%
							\pgf@y=\ht\pgfnodeparttextbox%
							\advance\pgf@y by\dp\pgfnodeparttextbox%
							\pgfmathsetlength\pgf@yc{\pgfkeysvalueof{/pgf/inner ysep}}%
							\advance\pgf@y by 2\pgf@yc%
							\pgfmathsetlength\pgf@yb{\pgfkeysvalueof{/pgf/minimum height}}%
							\ifdim\pgf@y<\pgf@yb%
							\pgf@y=\pgf@yb%
							\fi%
							\ifdim\pgf@x<\pgf@y%
							\pgf@x=\pgf@y%
							\else
							\pgf@y=\pgf@x%
							\fi
							\pgf@x=#1.5\pgf@x%
							\advance\pgf@x by.5\wd\pgfnodeparttextbox%
							\pgfmathsetlength\pgf@xa{\pgfkeysvalueof{/pgf/outer xsep}}%
							\advance\pgf@x by#1\pgf@xa%
							\pgf@y=#1.5\pgf@y%
							\advance\pgf@y by-.5\dp\pgfnodeparttextbox%
							\advance\pgf@y by.5\ht\pgfnodeparttextbox%
							\pgfmathsetlength\pgf@ya{\pgfkeysvalueof{/pgf/outer ysep}}%
							\advance\pgf@y by#1\pgf@ya%
						}
						\makeatother
						
						\pgfdeclareshape{square}{
							\savedanchor\northeast{\squarecorner{}}
							\savedanchor\southwest{\squarecorner{-}}
							
							\foreach \x in {east,west} \foreach \y in {north,mid,base,south} {
								\inheritanchor[from=rectangle]{\y\space\x}
							}
							\foreach \x in {east,west,north,mid,base,south,center,text} {
								\inheritanchor[from=rectangle]{\x}
							}
							\inheritanchorborder[from=rectangle]
							\inheritbackgroundpath[from=rectangle]
						}
						
						\tikzset{stretch/.initial=1}
						\newcommand\drawloop[4][]%
						{\draw[shorten <=0pt, shorten >=0pt,#1]
							($(#2)!\pgfkeysvalueof{/tikz/stretch}!(#2.#3)$)
							let \p1=($(#2.center)!\pgfkeysvalueof{/tikz/stretch}!(#2.north)-(#2)$),
							\n1= {veclen(\x1,\y1)*sin(0.5*(#4-#3))/sin(0.5*(180-#4+#3))}
							in arc [start angle={#3-90}, end angle={#4+90}, radius=\n1]%
						}

						\definecolor{cE8}{HTML}{0000FF}
						\definecolor{cE7}{HTML}{FF0000}
						\definecolor{cE6}{HTML}{FFA500}
						
						\tikzset{
							flavor/.style={rectangle, draw=black!100, thick, minimum size=2mm},
							gauge/.style={circle, thick, draw=black!100,fill=white!100,  minimum size=2mm, inner sep=0pt},
							bodyE6/.style={draw=cE6!100},
							bodyE7/.style={draw=cE7!100},
							bodyE8/.style={draw=cE8!100},
						}
						
						\crefname{table}{Table}{Tables}
						\crefname{figure}{Figure}{Figures}
						\crefname{section}{Section}{Sections}
						
						\hypersetup{
							pdftitle={From Regular to Irregular: A Unified Origin for Argyres--Douglas Theories},    
							pdfauthor={\textcopyright\ Simone Giacomelli, William Harding, Noppadol Mekareeya, Alessandro Mininno},     
							pdfsubject={HEP},   
							pdfcreator={pdfLaTex},   
							pdfproducer={LaTex}, 
							pdfkeywords={},
							colorlinks=true,
						}

						\title{From Regular to Irregular: A Unified Origin for Argyres--Douglas Theories}
						\author[a]{Simone Giacomelli\,\orcidlink{0000-0001-5873-0100},}
						\author[a,b]{William Harding\,\orcidlink{0009-0002-9842-433X},}
						\author[b,c]{Noppadol Mekareeya\,\orcidlink{0000-0002-3557-7489}}
						\author[d]{\\and Alessandro Mininno\,\orcidlink{0000-0002-9593-0440}}
						\affiliation[a]{Dipartimento di Fisica, Universit\`a di Milano-Bicocca, \\ Piazza della Scienza 3, I-20126 Milano, Italy}
						\affiliation[b]{INFN, sezione di Milano-Bicocca, \\Piazza della Scienza 3,  I-20126 Milano, Italy}
						\affiliation[c]{Department of Physics, Faculty of Science, Chulalongkorn University, \\ Phayathai Road,
							Pathumwan, Bangkok 10330, Thailand}
						\affiliation[d]{Department of Physics, University of Wisconsin--Madison,\\
							1150 University Avenue, Madison, WI 53706, USA}
						\emailAdd{simone.giacomelli@unimib.it}
						\emailAdd{w.harding@campus.unimib.it}
						\emailAdd{n.mekareeya@gmail.com}
						\emailAdd{mininno@physics.wisc.edu}
						\abstract{
							We propose that Argyres--Douglas theories of type $D_p(\mathrm{SU}(N))$ and $(A_{p-1}, A_{N-1})$ --- both realizable as Type A class $\mathcal{S}$ theories with irregular punctures --- can be obtained via a sequence of mass deformations from a common ancestor: a class $\mathcal{S}$ theory with only regular punctures. Building on our previous work, this result establishes that these theories ultimately originate from 6d $\mathcal{N}=(1,0)$ orbi-instanton theories compactified on a torus. The requisite 4d mass deformations are realized as tractable Fayet--Iliopoulos deformations on the 3d mirror quiver. The core of our method is a constructive procedure that utilizes the Euclidean algorithm to define a chain of deformations connecting different $D_p(\mathrm{SU}(N))$ theories. By reversing this chain, we recursively build a ``parent'' star-shaped quiver for any given $(N,p)$. This quiver is the 3d mirror theory of the required class $\mathcal{S}$ ancestor. We substantiate our general claims with several detailed examples that explicitly illustrate the deformation procedure.
						}

						\allowdisplaybreaks[1]
						
\begin{document} 
							
							\maketitle
							
							\section{Introduction and Summary}
							
							One of the largest known families of 4d $\mathcal{N}=2$ superconformal field theories (SCFTs) is obtained by compactifying 6d $\mathcal{N}=(2,0)$ theories on a Riemann surface with punctures. This framework, known as the class $\mathcal{S}$ construction \cite{Gaiotto:2009we,Gaiotto:2009gz}, has produced a rich landscape of 4d theories that have been studied extensively over the past fifteen years.\footnote{See, e.g., \cite{Akhond:2021xio,Argyres:2022mnu} for recent reviews.} In this paper, we consider class $\mathcal{S}$ theories of Type A with both regular and irregular punctures, focusing on the so-called Argyres--Douglas (AD) theories \cite{Argyres:1995xn,Argyres:1996hc,Eguchi:1996vu,Eguchi:1996ds,Shapere:1999xr,Cecotti:2010fi,Xie:2012hs}. These include $D_p(\SU(N))$ theories \cite{Cecotti:2013lda} (see also \cite{Cecotti:2012jx,Wang:2018gvb}) and$(A_{p-1}, A_{N-1})$ theories \cite{Cecotti:2010fi}.
							
							The $D_p(\SU(N))$ class is realized within the Type A class $\mathcal{S}$ framework by considering a Riemann sphere with one full regular puncture and one irregular (Type I) puncture \cite{Xie:2012hs,Xie:2013jc}. These models also admit a geometric engineering realization as a Type IIB string theory compactification on a non-compact Calabi--Yau (CY) threefold, realized as the zero-locus of a single hypersurface singularity in $\BC^3 \times \BC^*$ \cite{Cecotti:2010fi,Giacomelli:2017ckh}. As most of these theories are strongly coupled, an $\mathcal{N}=2$ Lagrangian description is not yet available. Nevertheless, many of their properties can be computed exactly via their geometric realization.
							
							Although the moduli space of $D_p(\SU(N))$ theories is now well understood, the Renormalization Group (RG) flows between them remain relatively unexplored, particularly those triggered by relevant deformations. RG flows initiated by Higgsing, for instance, are better understood due to the constraints of 't~Hooft anomaly matching. A key example is the flow from a $D_{p+N}(\SU(N))$ theory (with $p>0$) to the $(A_{p-1}, A_{N-1})$ theory by fully closing the regular puncture. This process involves assigning a vacuum expectation value (VEV) to the $\SU(N)$ moment map and constitutes a Higgs branch RG flow. The same $(A_{p-1}, A_{N-1})$ theory can also be reached from $D_p(\SU(N))$ via the Maruyoshi--Song (MS) flow \cite{Maruyoshi:2016aim,Maruyoshi:2016tqk,Agarwal:2016pjo,Giacomelli:2017ckh}.
							
							In recent years, many properties of the $D_p(\SU(N))$ theories (or more generally, $D_p^b(G)$ theories) have been investigated \cite{Giacomelli:2017ckh,DelZotto:2020esg,Closset:2020scj,Closset:2020afy,Bah:2021mzw, Bah:2021hei,Hosseini:2021ged,Kang:2021lic,Kang:2021ccs,Bhardwaj:2021mzl,Kang:2022zsl, Bah:2022yjf,Kang:2022vab,Beem:2023ofp,Couzens:2023kyf,Kang:2024inx}. Notably, in \cite{Giacomelli:2020ryy} (see also \cite{Xie:2012hs, DelZotto:2014kka}), 3d mirror theories were proposed in the form of $\cN=4$ Lagrangian theories represented by unitary quivers.\footnote{The 3d mirror theories for most of the $D_p^b(G)$ theories have been found in a series of works \cite{Carta:2021whq,Carta:2022spy,Carta:2022fxc,Carta:2023bqn}, containing also a detailed study of their conformal manifolds and generalized symmetries.} Generally, these theories possess a global symmetry algebra $\mathfrak{su}(N) \oplus \u(1)^{\mu-1}$, where $(\mu-1)$ is the number of additional mass parameters, excluding the Casimirs of $\SU(N)$.\footnote{This symmetry can get enhanced further to a larger Lie algebra that contains $\mathfrak{su}(N) \oplus \u(1)^{\mu-1}$ as a subalgebra.} The value $\mu$ corresponds to the greatest common divisor of $p$ and $N$.
							
							The purpose of this work is to build on our recent findings in \cite{Giacomelli:2024ycb}, where we demonstrated that all class $\mathcal{S}$ theories of Type A with regular untwisted punctures originate from mass deformations of orbi-instanton theories \cite{Aspinwall:1997ye,DelZotto:2014hpa,Heckman:2015bfa,Mekareeya:2017jgc}. These parent theories arise from the dimensional reduction on a two-torus of 6d $\mathcal{N}=(1,0)$ theories realized on M5-branes probing an M9-brane on a $\BC^2/\BZ_k$ singularity. Our central message is that all Type A class $\mathcal{S}$ theories with untwisted punctures can be recovered from 6d $\mathcal{N}=(1,0)$ theories on a torus, followed by relevant (mass) deformations in 4d. This approach obviates the need to consider Riemann surfaces of higher genus or with punctures at the 6d level.
							
							Since the resulting 4d theories are often strongly coupled, analyzing these mass deformations directly is a non-trivial task. We circumvent this by mapping the 4d relevant deformation to a Fayet--Iliopoulos (FI) deformation in the corresponding 3d mirror theory. The advantage is that the 3d mirror is a Lagrangian theory, where the effect of the FI term can be reliably studied via the equations of motion \cite{vanBeest:2021xyt,Bourget:2023uhe,Giacomelli:2024ycb}. We then posit that the infrared fixed point of the 4d RG flow is the theory whose 3d mirror is the quiver obtained after the FI deformation.
							
							In this work, we show that class $\mathcal{S}$ theories with \textit{irregular} punctures also adhere to this pattern. Specifically, we demonstrate that the $D_p(\SU(N))$ theories can be obtained by activating suitable mass deformations in class $\mathcal{S}$ theories that have only regular punctures. Furthermore, we find that a mass deformation of the $\SU(N)$ symmetry in a $D_p(\SU(N))$ theory induces an RG flow to the $(A_{p-1}, A_{N-1})$ AD theory. This flow shares its UV and IR fixed points with the MS flow, but manifestly preserves $\mathcal{N}=2$ supersymmetry.
							
							A central result of this work is the following:
							\begin{mdframed}[backgroundcolor=white, shadow=true, shadowsize=4pt,shadowcolor=black,
								roundcorner=6pt]
								Argyres--Douglas theories of type $D_p(\mathrm{SU}(N))$ and $(A_{p-1}, A_{N-1})$ share a unified origin. These theories, both characterized as Type A class $\mathcal{S}$ theories with irregular punctures, arise from a common ancestor --- a class $\mathcal{S}$ theory with only regular punctures --- via a sequence of mass deformations. Combined with our previous work \cite{Giacomelli:2024ycb}, this demonstrates that their ultimate origin lies in the torus compactification of 6d $\mathcal{N}=(1,0)$ orbi-instanton theories.
							\end{mdframed}

							\subsection*{Outlook}
							
							The $D_p(\SU(N))$ and $(A_m, A_n)$ theories represent some of the simplest instances of geometric engineering in Type IIB string theory. Our results suggest that torus compactifications (possibly with a twist along the cycles of the torus) of 6d theories, when supplemented with all possible relevant deformations in 4d, offer a general method for exploring the landscape of 4d SCFTs with eight supercharges, at least those realizable in string theory. It would therefore be of great interest to generalize our analysis to further test this claim by studying, for instance, other types of class $\mathcal{S}$ theories or other $D_p(G)$ and $(G, G')$ AD theories.

							\subsubsection*{Structure of the Paper}
							
							The paper is organized as follows. In Section~\ref{sec:FIDeformations-review}, we review the application of Fayet--Iliopoulos (FI) deformations to 3d $\mathcal{N}=4$ theories, based on the analyses in \cite{vanBeest:2021xyt,Bourget:2023uhe,Giacomelli:2024ycb}. In Section~\ref{sec:newrulesFI}, we introduce new FI deformation rules that are instrumental to our results.
							
							In Section~\ref{sec:Impclassdeform}, we study various classes of FI deformations applied to the 3d mirrors of $D_p(\SU(N))$ theories, which are crucial to derive our main result. We begin in Section~\ref{sec:DeformingDpSUNintoDpSUN-1} by showing that the 3d mirror of $D_\mu(\SU(\alpha\mu))$ can be deformed into the mirror of $D_\mu(\SU(\alpha\mu-1))$, providing a novel example of an FI deformation that transforms one star-shaped quiver into another corresponding to a $D_p(\SU(N))$ theory. Subsequently, in Section~\ref{sec:DeformingDpSUintoDNSU}, we demonstrate a key result: the mirror of a $D_p(\SU(N))$ theory with $p \leq N$ can be deformed to yield the mirror of $D_N(\SU(p))$. This deformation is a cornerstone of the general ansatz presented in Section \ref{sec:StarShapedforDpSUN}.
							
							In Section~\ref{sec:StarShapedforDpSUN}, we propose a general star-shaped quiver that, through specific deformations, flows to the 3d mirror of any $D_p(\SU(N))$ theory. We explain the rationale behind this ansatz and provide several instructive examples. Further examples and a general strategy for deforming these star-shaped quivers are detailed in Section~\ref{sec:examples}.
							Finally, Appendix~\ref{sec:DpSUmirror-review} provides a review of the construction of 3d mirror theories for all $D_p(\SU(N))$ Argyres--Douglas theories, while Appendix~\ref{sec:OItoStarShaped} reviews the procedure for obtaining any star-shaped quiver by deforming orbi-instanton theories.

							\subsubsection*{Conventions and Notation}
							
							\begin{enumerate}
								\item We primarily consider unitary quiver gauge theories. A circle \tikz[baseline=-0.5ex]{\node[gauge] {};} represents a gauge node, with a label indicating the rank of the group. A square \tikz[baseline=-0.5ex]{\node[flavor] {};} represents a flavor node, with its label denoting the number of fundamental hypermultiplets. A line connecting two nodes represents a hypermultiplet in the bifundamental representation. If a line corresponds to multiple hypermultiplets, this will be specified. A loop on a unitary gauge node indicates an adjoint-valued hypermultiplet.
								
								\item We adopt the concept of the ``excess number" ($\texttt{e}$), following the classification of gauge nodes as balanced, underbalanced, or overbalanced from \cite{Gaiotto:2008ak}. For a $\U(r)$ gauge group with $N_f$ fundamental hypermultiplets, the excess number is defined as:
								\begin{equation}
									\texttt{e}_{\U(r)} = N_f - 2r \fstop
								\end{equation}
								A quiver theory is deemed ``good" if all its gauge nodes have $\texttt{e}_{\U(r)} \geq 0$, ensuring that all monopole operators respect the unitarity bound. A node is \textit{balanced} if $\texttt{e}_{\U(r)}=0$ and \textit{overbalanced} if $\texttt{e}_{\U(r)} > 0$. Conversely, a node is \textit{underbalanced} if $\texttt{e}_{\U(r)} < 0$. A theory is called ``ugly" if any gauge group has $\texttt{e}_{\U(r)}=-1$, indicating a monopole operator saturating the unitarity bound. If any gauge group has $\texttt{e}_{\U(r)} < -1$, the theory is ``bad", as it contains monopole operators that violate the unitarity bound. It is important to note that a theory which appears ``good" may be revealed to be ``bad" when considering monopoles charged under multiple gauge groups, as is the case for certain affine Dynkin-shaped unitary quivers \cite{Cremonesi:2014xha}.\footnote{For a similar discussion applied to orthosymplectic magnetic quivers, see \cite{Lawrie:2024wan,Lawrie:2025exx}.}
								
								\item For a $T_\rho[\SU(N)]$ theory, if the partition $\rho$ is not specified (e.g., $T[\SU(N)]$), it is assumed to be the principal partition $\rho=\left[1^N\right]$. We may also drop the $[\SU(N)]$ when there is no ambiguity, referring to the theory simply as $T_\rho$.
								
								\item As detailed in Appendix~\ref{sec:DpSUmirror-review}, for $D_p(\SU(N))$ theories, we frequently use the definitions:
								\begin{equation}\label{eq:notationDpSU}
									\mu = \GCD(N,p) \coma \quad N = \mu \fn \coma \quad p = \mu \fp \fstop
								\end{equation}
								Furthermore, when discussing theories with $p \leq N$, we adopt the notation from \cite{Giacomelli:2020ryy}:
								\begin{equation}
									x = \left\lfloor \frac{N}{p} \right\rfloor \coma \quad M=N-p(x+1) \coma
								\end{equation}
								and
								\begin{equation}
									m_A = \fp(1+x)-\fn \coma \quad m_B = \fn-\fp x \coma \quad m_G = m_A m_B \fstop
								\end{equation}
								
								\item We use the term ``parent star-shaped quiver'' to refer to the 3d mirror of a 4d class $\mathcal{S}$ theory that serves as the parent theory for a given $D_p(\SU(N))$ theory, meaning that the latter can be obtained via a sequence of mass deformations starting from the class $\mathcal{S}$ theory in question. These mirrors are known to be star-shaped quivers according to \cite{Benini:2010uu}.
							\end{enumerate}

							\section{Rules of Fayet--Iliopoulos and Mass Deformations}
							\label{sec:FIDeformations-review}
							
							The application of Fayet--Iliopoulos deformations to 3d $\mathcal{N}=4$ theories was introduced in \cite{vanBeest:2021xyt} to study the Higgs branch of 5d SCFTs arising from circle compactifications of orbi-instanton theories. In that context, FI deformations on the magnetic quivers correspond to mass deformations on the 5d Higgs branch, allowing for an analysis of its structure when moving around the extended Coulomb branch. The set of possible deformations on magnetic quivers, first explored in \cite{vanBeest:2021xyt,Bourget:2023uhe}, was recently reviewed and extended in \cite{Giacomelli:2024ycb}. In this section, we briefly review the established rules for FI deformations on unitary magnetic quivers and introduce new deformations that are essential for the present work.
							
							\subsection{Review of Basic Rules}
							\label{sec:FIdefooldrules}
							
							Let us first review the action of an FI parameter $\xi$ on the following quiver:
							\begin{equation}
								\begin{tikzpicture}[baseline,font=\footnotesize]
									\node[gauge,label=below:{$N$}] (Nu) {};
									\node[flavor,label=below:{$k$}] (ku) [right=12mm of Nu] {};
									\draw[thick] (Nu) to[out=135, in=215, looseness=12] node[left,pos=0.5] {$\Phi$} (Nu);
									\draw[thick] (Nu) -- node[above,pos=0.5] {$\widetilde{Q}_i,Q_i$} (ku);
								\end{tikzpicture}
							\end{equation}
							where $\Phi$ is the adjoint chiral in the $\mathcal{N}=4$ vector multiplet. The superpotential acquires an additional term, namely
							\begin{equation} 
								\mathcal{W}= \widetilde{Q}_i\Phi Q^i + \xi\Tr\Phi\coma
							\end{equation}
							which modifies the F-term and D-term equations to become
							\begin{equation}
								\begin{split}
									Q^i\widetilde{Q}_i &= \xi I_N\coma\\ 
									Q^iQ_i^{\dagger}-\widetilde{Q}^{\dagger i} \widetilde{Q}_i &=0\fstop
								\end{split}
							\end{equation}
							The sum over flavor indices includes all bifundamental hypermultiplets charged under the $\U(N)$ gauge node. For a generic quiver composed of multiple $\U(N_i)$ gauge groups and only bifundamental matter, this implies that the FI parameters must satisfy the constraint
							\be\label{constr} \sum_i N_i\xi_i=0\fstop\ee
							This condition requires that at least two gauge nodes must have a non-zero FI parameter turned on. Another crucial property is that FI deformations preserve the dimension of the Higgs branch of the quiver.
							
							\subsubsection{Fayet--Iliopoulos Deformations at Nodes of the Same Rank}
							
							The simplest FI deformation involves activating FI parameters for two nodes of the same rank, say $\U(k)$. Solving the F-term and D-term equations results in the breaking of all $\U(n_i)$ gauge groups to $\U(n_i-k)$ along the subquiver connecting the two $\U(k)$ nodes, while all other nodes remain unaffected. Among the broken $\U(k)$ factors, a diagonal combination survives, leading to the addition of a new $\U(k)$ node. This new node is attached to all nodes in the original quiver whose excess number was altered by the deformation, thereby restoring the original balance of the quiver. The multiplicity of the edges connecting this new $\U(k)$ node is determined by the number of flavors required to restore the original excess numbers. This operation can be viewed as a modified quiver subtraction, where the subtracted part consists only of $\U(k)$ nodes and the balance is restored by a new $\U(k)$ node.
							
							Another deformation, analyzed in \cite{Giacomelli:2024ycb}, involves activating FI parameters at nodes with ranks $\U(n)$, $\U(k)$, and $\U(n+k)$ such that $\xi_n=\xi_k$. This can be described as a two-step process. First, a subquiver between the $\U(n)$ and $\U(n+k)$ nodes is subtracted. After rebalancing, a further subtraction is performed on the resulting quiver between the $\U(k)$ node and the original $\U(n+k)$ node (which became $\U(k)$ after the first step). When performing the deformation in two steps, it is important not to rebalance the rebalancing $\U(n)$ node from the first step with the $\U(k)$ node from the second. For an explicit example, we refer the reader to \cite[Section 3.2]{Giacomelli:2024ycb}.
							
							\subsection{New Fayet--Iliopoulos Deformations}
							\label{sec:newrulesFI}
							
							Let us consider the following quiver, in which FI parameters have been activated at the {\color{red}{red}} nodes:
							\begin{equation}\label{Ftailmove1}
								\begin{tikzpicture}[baseline=0,font=\footnotesize]
									\node[gauge,label=below:{$1$},fill=red] (uB1) {};
									\node[gauge,label=below:{$2$}] (uB2) [right=6mm of uB1] {};
									\node (dotsB1) [right=6mm of uB2] {$\cdots$};
									\node[gauge,label=below:{$r$}] (uBp) [right=6mm of dotsB1] {};
									\node[gauge,label=below:{$1$},fill=red] (u1) [right=10mm of uBp] {};
									\draw[thick] (uB1) -- (uB2) -- (dotsB1) -- (uBp) -- node[above,midway] {$h+1$} (u1);
								\end{tikzpicture}
							\end{equation}
							where we assume that $r\leq h$ so that the whole quiver is ``good". After the deformation, we obtain the quiver
							\begin{equation}\label{Ftailmove2}
								\begin{tikzpicture}[baseline=0,font=\footnotesize]
									\node[gauge,label=below:{$1$},fill=red] (uB1) {};
									\node[gauge,label=below:{$2$}] (uB2) [right=6mm of uB1] {};
									\node (dotsB1) [right=6mm of uB2] {$\cdots$};
									\node[gauge,label=below:{$r-1$}] (uBp) [right=6mm of dotsB1] {};
									\node[gauge,label=below:{$1$},fill=red] (u1) [right=10mm of uBp] {};
									\draw[thick] (uB1) -- (uB2) -- (dotsB1) -- (uBp) -- node[above,midway] {$h$} (u1);
									\node [below=8mm of uB1, xshift=21mm] {$+\,h$ free hypermultiplets.};
								\end{tikzpicture}
							\end{equation}
							The transition from \eqref{Ftailmove1} to \eqref{Ftailmove2} can be understood as follows. The FI deformation causes all bifundamental fields along the tail to acquire a VEV. Modulo an $\SU(h+1)$ transformation, we can assume that only one of the $\U(r)\times \U(1)$ bifundamentals gets a VEV. Its components are removed from the spectrum as they pair up with vector multiplets or otherwise become massive. Consequently, all gauge groups along the tail decrease in rank by one unit. In particular, the $\U(r)$ node is Higgsed to $\U(r-1)$. The remaining $h$ bifundamentals between the $\U(r)$ node and the $\U(1)$ node become an equal number of $\U(r-1)\times \U(1)$ bifundamentals, plus $h$ singlets that decouple as free hypermultiplets. This leads to the final state in \eqref{Ftailmove2}.\footnote{In order to avoid potential confusion, we emphasize again that, in the transition from \eqref{Ftailmove1} to \eqref{Ftailmove2}, subtracting the abelian quiver reduces the rank of each node by one. This removes the two flanking {\color{red}{red}} $\U(1)$ nodes in \eqref{Ftailmove1} and maps the remaining $\U(x)$ nodes to $\U(x-1)$. In particular, the $\U(2)$ node in \eqref{Ftailmove1} becomes the $\U(1)$ node \eqref{Ftailmove2}, and so on. The new rightmost {\color{red}{red}} node in \eqref{Ftailmove2}, along with edge multiplicity $h$, is then introduced via rebalancing.}
							
							This process can be rephrased as a quiver subtraction where the subtracted part is an abelian quiver with $r+1$ nodes, removing the $h+1$ bifundamentals in the process. A rebalancing $\U(1)$ node is then added, connected only to the $\U(r-1)$ node with multiplicity $h$ to preserve the excess number. Since FI deformations preserve the Higgs branch dimension, the initial and final quivers must have Higgs branches of equal dimension, which holds if the $h$ free hypermultiplets in \eqref{Ftailmove2} are included.
							
							An alternative argument, which is helpful for later sections, proceeds as follows. Let us start with a full tail attached to a bouquet of $h+1$ abelian nodes. For simplicity, we consider the case where one node in the bouquet is connected by a single bifundamental to the other $h$ nodes:
							
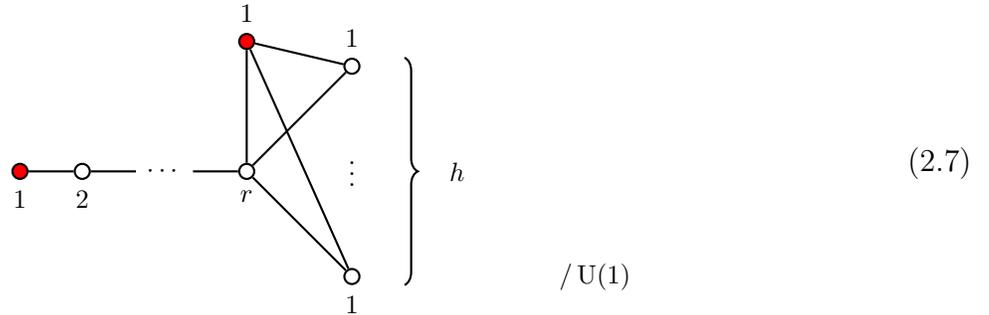
\begin{equation}\label{Fanmove1}
								\begin{tikzpicture}[baseline=0,font=\footnotesize]
									\node[gauge,label=below:{$r$}] (uN1) {};
									\node (dots) [left=6mm of uN1] {$\cdots$};
									\node[gauge,label=below:{$2$}] (u2) [left=6mm of dots] {};
									\node[gauge,label=below:{$1$},fill=red] (u1) [left=6mm of u2]{};
									\node[gauge,label=above:{$1$}] (p2) [above right=1.75cm of uN1] {};
									\node[gauge,label=below:{$1$}] (p5) [below right=1.75cm of uN1] {};
									\node (dotsp2) [below=8mm of p2] {$\vdots$};
									\node (U1d) [right=25mm of p5] {$/\U(1)$};
									\node[gauge,label=above:{$1$},fill=red] (p3) [above=1.5cm of uN1] {};
									\draw[thick] (u1) -- (u2) -- (dots) -- (uN1);
									\draw[thick] (uN1) --  (p2);
									\draw[thick] (uN1) --  (p3);
									\draw[thick] (uN1) --  (p5);
									\draw[thick] (p3) --  (p2);
									\draw[thick] (p3) --  (p5);
									\draw[thick] [decorate,decoration={brace,amplitude=5pt},xshift=0cm, yshift=0cm]
									([xshift=7mm]p2.north) -- ([xshift=7mm]p5.south) node [black,pos=0.5,xshift=0.7cm] 
									{$h$};
								\end{tikzpicture}
							\end{equation}
							Here, $/\U(1)$ denotes the overall $\U(1)$ that decouples. We first perform an FI deformation involving the {\color{red}{red}} nodes, which leads to
							
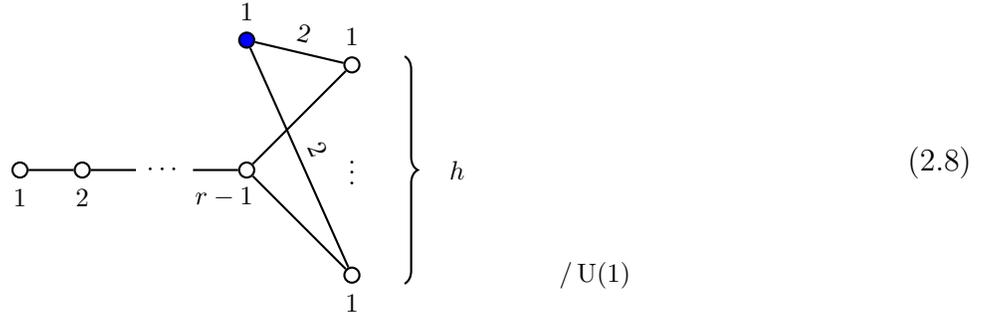
\begin{equation}\label{Fanmove2-new}
								\begin{tikzpicture}[baseline=0,font=\footnotesize]
									\node[gauge,label={[xshift=-0.3cm,yshift=-0.75cm]:{$r-1$}}] (uN1) {};
									\node (dots) [left=6mm of uN1] {$\cdots$};
									\node[gauge,label=below:{$2$}] (u2) [left=6mm of dots] {};
									\node[gauge,label=below:{$1$}] (u1) [left=6mm of u2]{};
									\node[gauge,label=above:{$1$}] (p2) [above right=1.75cm of uN1] {};
									\node[gauge,label=below:{$1$}] (p5) [below right=1.75cm of uN1] {};
									\node (dotsp2) [below=8mm of p2] {$\vdots$};
									\node (U1d) [right=25mm of p5] {$/\U(1)$};
									\node[gauge,label=above:{$1$},fill=blue] (p3) [above=1.5cm of uN1] {};
									\draw[thick] (u1) -- (u2) -- (dots) -- (uN1);
									\draw[thick] (uN1) --  (p2);
									\draw[thick] (uN1) --  (p5);
									\draw[thick] (p3) -- node[midway,above,sloped] {$2$} (p2);
									\draw[thick] (p3) -- node[midway,above,sloped] {$2$} (p5);
									\draw[thick] [decorate,decoration={brace,amplitude=5pt},xshift=0cm, yshift=0cm]
									([xshift=7mm]p2.north) -- ([xshift=7mm]p5.south) node [black,pos=0.5,xshift=0.7cm] 
									{$h$};
								\end{tikzpicture}
							\end{equation}
							If we now turn on FI parameters for the {\color{blue}blue} node and another $\U(1)$ node from the remaining $h$ nodes, we obtain
							
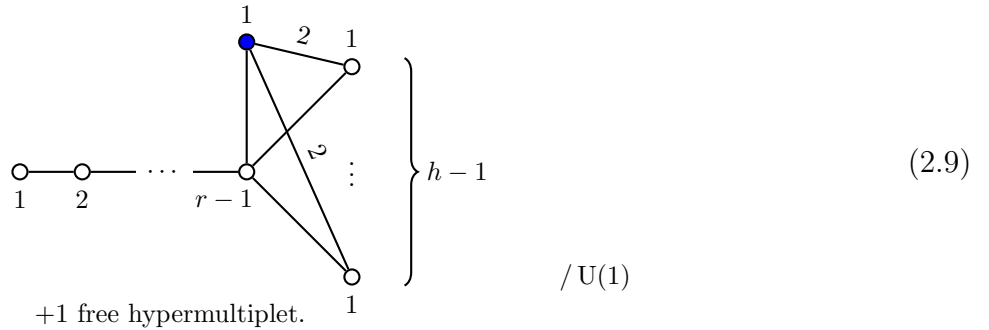
\begin{equation}\label{Fanmove3-new}
								\begin{tikzpicture}[baseline=0,font=\footnotesize]
									\node[gauge,label={[xshift=-0.3cm,yshift=-0.75cm]:{$r-1$}}] (uN1) {};
									\node (dots) [left=6mm of uN1] {$\cdots$};
									\node[gauge,label=below:{$2$}] (u2) [left=6mm of dots] {};
									\node[gauge,label=below:{$1$}] (u1) [left=6mm of u2]{};
									\node[gauge,label=above:{$1$}] (p2) [above right=1.75cm of uN1] {};
									\node[gauge,label=below:{$1$}] (p5) [below right=1.75cm of uN1] {};
									\node (dotsp2) [below=8mm of p2] {$\vdots$};
									\node (U1d) [right=25mm of p5] {$/\U(1)$};
									\node[gauge,label=above:{$1$},fill=blue] (p3) [above=1.5cm of uN1] {};
									\draw[thick] (u1) -- (u2) -- (dots) -- (uN1);
									\draw[thick] (uN1) --  (p2);
									\draw[thick] (uN1) --  (p3);
									\draw[thick] (uN1) --  (p5);
									\draw[thick] (p3) -- node[midway,above,sloped] {$2$} (p2);
									\draw[thick] (p3) -- node[midway,above,sloped] {$2$} (p5);
									\draw[thick] [decorate,decoration={brace,amplitude=5pt},xshift=0cm, yshift=0cm]
									([xshift=7mm]p2.north) -- ([xshift=7mm]p5.south) node [black,pos=0.5,xshift=0.7cm] 
									{$h-1$};
									\node [below=15mm of u1, xshift=20mm] {$+1$ free hypermultiplet.};
								\end{tikzpicture}
							\end{equation}
							Reiterating this move other $h-1$ times, we precisely recover the quiver in \eqref{Ftailmove2}. We expect to obtain the same result if we proceed in reverse order: first collapsing $h$ of the $\U(1)$ nodes in \eqref{Fanmove1} to arrive at \eqref{Ftailmove1}, and then performing the transition from \eqref{Ftailmove1} to \eqref{Ftailmove2} as desired.

							\section{Important Classes of Deformations}
							\label{sec:Impclassdeform}
							In this section, we analyze several important classes of deformations that are instrumental in deriving our main result in the subsequent section.
							
							\subsection{Deforming \texorpdfstring{$D_\mu(\SU(\alpha \mu))$}{Dmu(SU(alpha mu))} to \texorpdfstring{$D_\mu(\SU(\alpha\mu-1))$}{Dmu(SU(alpha mu - 1))}}
							\label{sec:DeformingDpSUNintoDpSUN-1}
							
							Here, we show that $D_\mu(\SU(\alpha \mu-1))$ can be obtained by a series of mass deformations from $D_\mu(\SU(\alpha \mu))$. This statement is equivalent to the claim that the star-shaped parent quiver for $D_\mu(\SU(\alpha \mu-1))$ is identical to that for $D_\mu(\SU(\alpha \mu))$. Note that $D_\mu(\SU(\alpha \mu))$ is a Lagrangian theory described by
							\bes{
								D_\mu(\SU(\alpha \mu)): \qquad \SU(\alpha)-\SU(2\alpha)-\SU(3\alpha)-\cdots - \SU(\alpha(\mu-1))-[\alpha \mu]\fstop
							}
							Its 3d mirror theory (see Appendix \ref{sec:DpSUNmirrorsp<=N}) is the following star-shaped quiver:
							
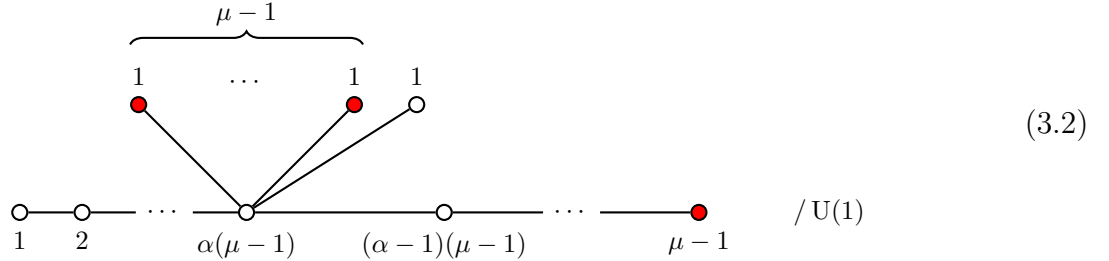
\begin{equation}\label{eq:generalDmuSUalphamu}
								\begin{tikzpicture}[baseline=30,font=\footnotesize]
									\node[gauge,label=below:{$\alpha(\mu-1)$}] (uBM) {};
									\node (dotsB1) [left=6mm of uBM] {$\cdots$};
									\node[gauge,label=below:{$2$}] (uB2) [left=6mm of dotsB1] {}; 
									\node[gauge,label=below:{$1$}] (uB1) [left=6mm of uB2] {}; 
									\node[gauge,label=below:{$(\alpha -1)(\mu -1)$}] (uN1) [right=24mm of uBM]{};
									\node (dotsA1) [right=12mm of uN1] {$\cdots$};
									\node[gauge,label=below:{$\mu-1$},fill=red] (uNp) [right=12mm of dotsA1]{};
									\node (U1d) [right=10mm of uNp] {$/\U(1)$};
									\node[gauge,label=above:{$1$},fill=red] (u1) [above left=18mm of uBM] {};
									\node[gauge,label=above:{$1$},fill=red] (u12) [above right=18mm of uBM] {};
									\node[gauge,label=above:{$1$}] (u13) [right=6mm of u12] {};
									\node (dotsu1) [above=14mm of uBM] {$\cdots$};
									\draw[thick] [decorate,decoration={brace,amplitude=5pt},xshift=0cm, yshift=0cm]
									([yshift=8mm]u1.west) -- ([yshift=8mm]u12.east) node [black,pos=0.5,yshift=0.4cm] 
									{$\mu-1$};
									\draw[thick] (uBM) -- (dotsB1) -- (uB2) -- (uB1);
									\draw[thick] (uBM) -- (uN1) -- (dotsA1) -- (uNp);
									\draw[thick] (uBM) -- (u1);
									\draw[thick] (uBM) -- (u12);
									\draw[thick] (uBM) -- (u13);
								\end{tikzpicture}
							\end{equation}
							We claim that this quiver is, in fact, the star-shaped parent theory for $D_\mu(\SU(\alpha \mu-1))$.
							
							For convenience, following the convention introduced in \cite{Giacomelli:2020ryy} and reviewed in Appendix \ref{sec:DpSUNmirrorsp<=N}, we refer to the right tail consisting of the nodes $\U(\alpha(\mu-1))$, $\U((\alpha-1)(\mu-1))$, $\cdots$, $\U(\mu-1)$ as {\it Tail A}, while the left tail consisting of the nodes $\U(1)$, $\U(2)$, $\cdots$, $\U(\alpha(\mu-1))$ will be called {\it Tail B}. Finally, in the following, the $\U(\alpha(\mu-1))$ gauge node will be called {\it central node}. 
							
							Let us activate FI parameters for the $\mu-1$ $\U(1)$ nodes in the bouquet and the $\U(\mu-1)$ node of Tail A (the {\color{red}red} nodes). After the deformation, the quiver becomes
							
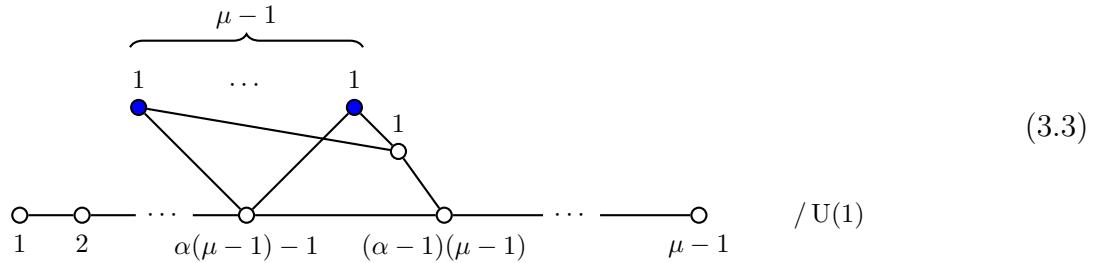
\begin{equation}\label{eq:generalDmuSUalphamu-2}
								\begin{tikzpicture}[baseline=30,font=\footnotesize]
									\node[gauge,label=below:{$\alpha(\mu-1)-1$}] (uBM) {};
									\node (dotsB1) [left=6mm of uBM] {$\cdots$};
									\node[gauge,label=below:{$2$}] (uB2) [left=6mm of dotsB1] {};
									\node[gauge,label=below:{$1$}] (uB1) [left=6mm of uB2] {};
									\node[gauge,label=below:{$(\alpha - 1)(\mu -1)$}] (uN1) [right=24mm of uBM]{};
									\node (dotsA1) [right=12mm of uN1] {$\cdots$};
									\node[gauge,label=below:{$\mu-1$}] (uNp) [right=12mm of dotsA1]{};
									\node (U1d) [right=10mm of uNp] {$/\U(1)$};
									\node[gauge,label=above:{$1$},fill=blue] (u1) [above left=18mm of uBM] {};
									\node[gauge,label=above:{$1$},fill=blue] (u12) [above right=18mm of uBM] {};
									\node[gauge,label=above:{$1$}] (u13) [below right=6mm of u12] {};
									\node (dotsu1) [above=14mm of uBM] {$\cdots$};
									\draw[thick] [decorate,decoration={brace,amplitude=5pt},xshift=0cm, yshift=0cm]
									([yshift=8mm]u1.west) -- ([yshift=8mm]u12.east) node [black,pos=0.5,yshift=0.4cm] 
									{$\mu-1$};
									\draw[thick] (uBM) -- (dotsB1) -- (uB2) -- (uB1);
									\draw[thick] (uBM) -- (uN1) -- (dotsA1) -- (uNp);
									\draw[thick] (uBM) -- (u1);
									\draw[thick] (uBM) -- (u12);
									\draw[thick] (uN1) -- (u13);
									\draw[thick] (u1) -- (u13);
									\draw[thick] (u12) -- (u13);
								\end{tikzpicture}
							\end{equation}
							where we introduced $\mu-1$ {\color{blue}{blue}} $\U(1)$ nodes to rebalance the quiver after the deformations. In order to obtain the 3d mirror of $D_\mu(\SU(\alpha \mu-1))$ from \eqref{eq:generalDmuSUalphamu-2}, we need to deform the bouquet of $\mu-1$ $\U(1)$ nodes and reduce them to a single $\U(1)$, since Tail A and Tail B of $D_\mu(\SU(\alpha \mu-1))$ are already present with
							\begin{equation}
								M = \alpha(\mu-1)-1\fstop
							\end{equation}
							To deform the bouquet of $\U(1)$s, we depict \eqref{eq:generalDmuSUalphamu-2} as follows:
							
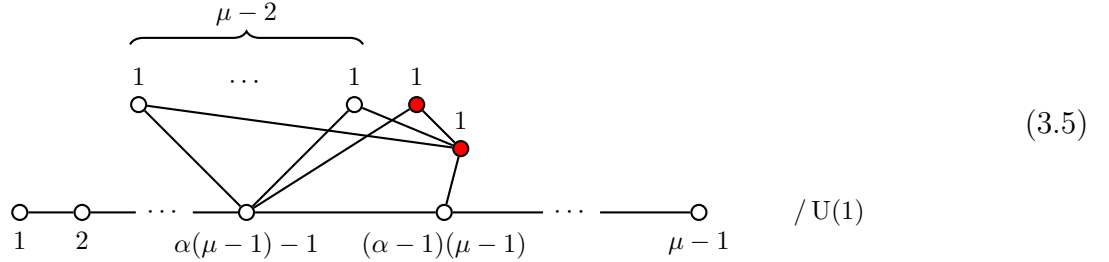
\begin{equation}\label{eq:generalDmuSUalphamu-3}
								\begin{tikzpicture}[baseline=30,font=\footnotesize]
									\node[gauge,label=below:{$\alpha(\mu-1)-1$}] (uBM) {};
									\node (dotsB1) [left=6mm of uBM] {$\cdots$};
									\node[gauge,label=below:{$2$}] (uB2) [left=6mm of dotsB1] {};
									\node[gauge,label=below:{$1$}] (uB1) [left=6mm of uB2] {};
									\node[gauge,label=below:{$(\alpha - 1)(\mu -1)$}] (uN1) [right=24mm of uBM]{};
									\node (dotsA1) [right=12mm of uN1] {$\cdots$};
									\node[gauge,label=below:{$\mu-1$}] (uNp) [right=12mm of dotsA1]{};
									\node (U1d) [right=10mm of uNp] {$/\U(1)$};
									\node[gauge,label=above:{$1$}] (u1) [above left=18mm of uBM] {};
									\node[gauge,label=above:{$1$}] (u12) [above right=18mm of uBM] {};
									\node[gauge,label=above:{$1$},fill=red] (u14) [right=6mm of u12] {};
									\node[gauge,label=above:{$1$},fill=red] (u13) [below right=6mm of u14] {};
									\node (dotsu1) [above=14mm of uBM] {$\cdots$};
									\draw[thick] [decorate,decoration={brace,amplitude=5pt},xshift=0cm, yshift=0cm]
									([yshift=8mm]u1.west) -- ([yshift=8mm]u12.east) node [black,pos=0.5,yshift=0.4cm] 
									{$\mu-2$};
									\draw[thick] (uBM) -- (dotsB1) -- (uB2) -- (uB1);
									\draw[thick] (uBM) -- (uN1) -- (dotsA1) -- (uNp);
									\draw[thick] (uBM) -- (u1);
									\draw[thick] (uBM) -- (u12);
									\draw[thick] (uBM) -- (u14);
									\draw[thick] (uN1) -- (u13);
									\draw[thick] (u1) -- (u13);
									\draw[thick] (u12) -- (u13);
									\draw[thick] (u14) -- (u13);
								\end{tikzpicture}
							\end{equation}
							where we have turned on FI parameters at the {\color{red} red} nodes. The deformation is implemented by subtracting an $A_2$ quiver connecting the two nodes, which results in
							
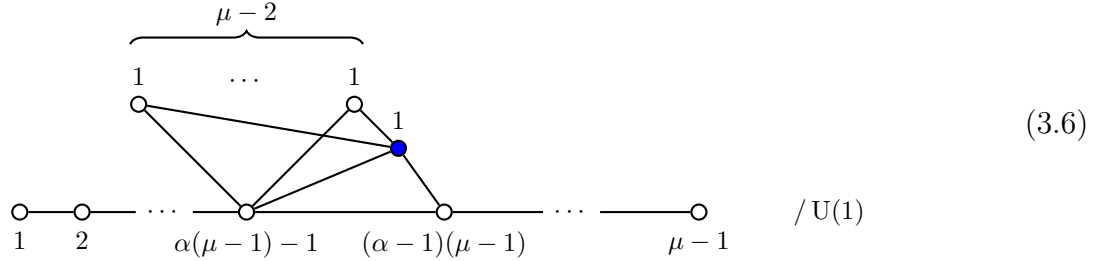
\begin{equation}\label{eq:generalDmuSUalphamu-4}
								\begin{tikzpicture}[baseline=30,font=\footnotesize]
									\node[gauge,label=below:{$\alpha(\mu-1)-1$}] (uBM) {};
									\node (dotsB1) [left=6mm of uBM] {$\cdots$};
									\node[gauge,label=below:{$2$}] (uB2) [left=6mm of dotsB1] {};
									\node[gauge,label=below:{$1$}] (uB1) [left=6mm of uB2] {};
									\node[gauge,label=below:{$(\alpha - 1)(\mu -1)$}] (uN1) [right=24mm of uBM]{};
									\node (dotsA1) [right=12mm of uN1] {$\cdots$};
									\node[gauge,label=below:{$\mu-1$}] (uNp) [right=12mm of dotsA1]{};
									\node (U1d) [right=10mm of uNp] {$/\U(1)$};
									\node[gauge,label=above:{$1$}] (u1) [above left=18mm of uBM] {};
									\node[gauge,label=above:{$1$}] (u12) [above right=18mm of uBM] {};
									\node[gauge,label=above:{$1$},fill=blue] (u13) [below right=6mm of u12] {};
									\node (dotsu1) [above=14mm of uBM] {$\cdots$};
									\draw[thick] [decorate,decoration={brace,amplitude=5pt},xshift=0cm, yshift=0cm]
									([yshift=8mm]u1.west) -- ([yshift=8mm]u12.east) node [black,pos=0.5,yshift=0.4cm] 
									{$\mu-2$};
									\draw[thick] (uBM) -- (dotsB1) -- (uB2) -- (uB1);
									\draw[thick] (uBM) -- (uN1) -- (dotsA1) -- (uNp);
									\draw[thick] (uBM) -- (u1);
									\draw[thick] (uBM) -- (u12);
									\draw[thick] (uBM) -- (u13);
									\draw[thick] (uN1) -- (u13);
									\draw[thick] (u1) -- (u13);
									\draw[thick] (u12) -- (u13);
								\end{tikzpicture}
							\end{equation}
							We can iterate the deformation for all the remaining $\U(1)$s in the bouquet, eventually obtaining
							
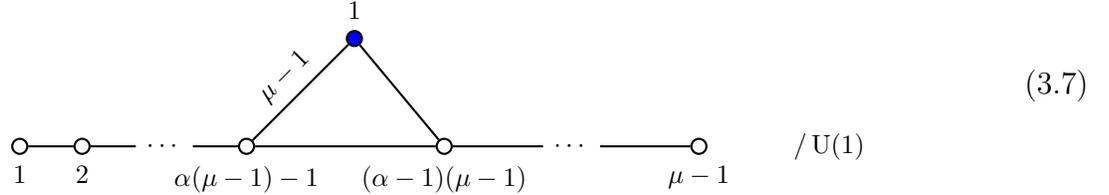
\begin{equation}
								\begin{tikzpicture}[baseline=20,font=\footnotesize]
									\node[gauge,label=below:{$\alpha(\mu-1)-1$}] (uBM) {};
									\node (dotsB1) [left=6mm of uBM] {$\cdots$};
									\node[gauge,label=below:{$2$}] (uB2) [left=6mm of dotsB1] {};
									\node[gauge,label=below:{$1$}] (uB1) [left=6mm of uB2] {};
									\node[gauge,label=below:{$(\alpha - 1)(\mu -1)$}] (uN1) [right=24mm of uBM]{};
									\node (dotsA1) [right=12mm of uN1] {$\cdots$};
									\node[gauge,label=below:{$\mu-1$}] (uNp) [right=12mm of dotsA1]{};
									\node (U1d) [right=10mm of uNp] {$/\U(1)$};
									\node[gauge,label=above:{$1$},fill=blue] (u12) [above right=18mm of uBM] {};
									\draw[thick] (uBM) -- (dotsB1) -- (uB2) -- (uB1);
									\draw[thick] (uBM) -- (uN1) -- (dotsA1) -- (uNp);
									\draw[thick] (uBM) --node[pos=0.5,above,sloped] {$\mu-1$} (u12);
									\draw[thick] (uN1) --  (u12);
								\end{tikzpicture}
							\end{equation}
							which is the 3d mirror of $D_\mu(\SU(\alpha\mu-1))$.   
							
							Let us summarize this process, which involves a sequence of FI deformations. For instance, consider two $\U(1)$ nodes in the bouquet connected to the central node and the rebalancing node. Activating FI parameters for these two nodes and the rebalancing node leads to a reduction in the number of nodes in the bouquet. Iterating this procedure collapses the entire bouquet into a single $\U(1)$ node connected with multiplicity $\mu-1$ to the central node, and also connected to Tail A. The resulting quiver is precisely the 3d mirror of $D_\mu(\SU(\alpha\mu-1))$.
							
							Note that the special case of $\mu=2$ can be inferred directly in 4d. This corresponds to deforming the $D_2(\SU(2n))$ theory—which is simply 4d $\mathcal{N}=2$ $\SU(n)$ SQCD with $2n$ flavors—to the $D_2(\SU(2n-1))$ theory. As pointed out in \cite[Section 7.2.1]{Cecotti:2013lda}, the latter can be realized at the maximally singular point of $\SU(n)$ SQCD with $2n-1$ flavors.
							
							\subsection{Deforming \texorpdfstring{$D_p(\SU(N))$}{Dp(SU(N))} with \texorpdfstring{$p\leq N$}{p<=N} to \texorpdfstring{$D_N(\SU(p))$}{DN(SU(p))}}
							\label{sec:DeformingDpSUintoDNSU}
							
							The mirror theory for $D_p(\SU(N))$ with $p \leq N$, as described in Appendix~\ref{sec:DpSUNmirrorsp<=N}, can be recursively deformed into the mirror of $D_N(\SU(p))$, which is detailed in Appendix~\ref{sec:DpSUNmirrorsp>=N} under the replacement $N\leftrightarrow p$. In this section, we explain how these deformations are performed. 
							
							We begin with the general mirror for $D_p(\SU(N))$, where we activate FI parameters at the {\color{red} red} nodes:
							\begin{equation}

							\end{equation}
							where $H_\text{\tiny free}$ is the number of free hypermultiplets, given in \eqref{eq:freehypers-p<=N}. Subsequently, we use the terminologies of Tail A and Tail B as defined in \eqref{def:TailATailB}. The resulting quiver, after rebalancing, is
							\begin{equation}
								%
							\end{equation}
							The process can be iterated $x$ times by switching on FI deformation at the $\U(p-1)$ nodes of Tails A and B, until Tail A disappears. In particular, Tail B reduces to a full tail with final node $\U(M-x(p-1))= \U(\mu m_B-1)$, and the resulting intermediate quiver is
							\begin{equation}
								%
								\label{eq:intermidiateDpSUmirror}
							\end{equation}
							We can apply the new rules described in Section \ref{sec:newrulesFI} to the {\color{red} red} nodes in \eqref{eq:intermidiateDpSUmirror}, obtaining the following quiver:
							\begin{equation}
								%
								\label{eq:intermidiateDpSUmirror-3}
							\end{equation}
							It is possible to iterate the procedure $m_B$ times turning on FI deformations at the {\color{red} red} $\U(1)$ node of the tail on the right and the balancing $\U(1)$ node in {\color{blue}{blue}}. The multiplicity of the bifundamental connecting the {\color{blue}{blue}} node to the tail will decrease until the edge completely disappears, while the multiplicity of the edges between the {\color{blue}{blue}} node and the other $\U(1)$s in the complete graph will become $m_G+m_B^2=\mathfrak{p}m_B$, i.e.
							\begin{equation}
								%
								\label{eq:intermidiateDpSUmirror-4}
							\end{equation}
							The process can also be repeated recursively for all $\mu-1$ $\U(1)$s left in the complete graph, so that tail B is completely removed. The final quiver is 
							\begin{equation}
								%
								\label{eq:intermidiateDpSUmirror-5}
							\end{equation}
							where we have already isolated one of the $x$ tails on the left and turned on the FI deformations that we will use next. The iterations are analogous to the previous one, and a tail will increase the multiplicity of the complete graph by $\mathfrak{p}^2$, so that we end up with
							\begin{equation}
								%
								\label{eq:intermidiateDpSUmirror-6}
							\end{equation}
							Repeating it $(x-1)$ times, we obtain
							\begin{equation}
								%
								\label{eq:mirrorDNSUp}
							\end{equation}
							where
							\begin{equation}
								\mathfrak{p}m_B+(x-1)\mathfrak{p}^2 = \mathfrak{p}(\mathfrak{n}-\mathfrak{p})\coma
							\end{equation}
							while 
							\begin{equation}
								\begin{split}
									\tilde{H}_\text{\tiny free}&=H_\text{\tiny free}+\frac{1}{2}\mu m_B(m_B-1)+\frac{1}{2}(x-1)\mu \mathfrak{p}(\mathfrak{p}-1)\\
									&=\frac{1}{2}\mu(\mathfrak{p}-1)(\mathfrak{n}-\mathfrak{p}-1)\fstop
								\end{split}
								\label{eq:HfreeDNSUp}
							\end{equation}
							The quiver in \eqref{eq:mirrorDNSUp}, together with the $\tilde{H}_\text{\tiny free}$ free hypermultiplets from \eqref{eq:HfreeDNSUp}, is precisely the mirror of $D_N(\SU(p))$ as described in \eqref{eq:generalDpSU-p>=N} (with the replacement $N\leftrightarrow p$).
							
							Finally, we note that the results of Section~\ref{sec:DeformingDpSUNintoDpSUN-1} imply that the theories $D_\mu(\SU(\alpha\mu-1))$, $D_{\alpha\mu-1}(\SU(\mu))$, and $D_{\alpha\mu}(\SU(\mu))$ can all be obtained from the star-shaped quiver in \eqref{eq:generalDmuSUalphamu}, which is the 3d mirror of $D_\mu(\SU(\alpha\mu))$.
							
							\subsection{Deforming \texorpdfstring{$D_N(\SU(p))$}{DN(SU(p))} to \texorpdfstring{$(A_{N-1},A_{p-1})$}{(A(N-1),A(p-1))}}
							\label{sec:DeformingANAp}
							
							We also find that, by continuing the procedure described above and deforming the final tail, we obtain a complete graph of $\U(1)$ nodes where each edge has multiplicity $\mathfrak{n}\mathfrak{p}$, along with 
							\begin{equation}
								\begin{split}
									H_\text{\tiny free}&=\frac{1}{2}\mu(\mathfrak{p}-1)(\mathfrak{n}-1)
								\end{split}
								\label{eq:HfreeAA}
							\end{equation}
							free hypermultiplets. This is precisely the mirror theory of $(A_{N-1},A_{p-1})$ \cite{DelZotto:2014kka,Giacomelli:2020ryy,Carta:2021whq}. We therefore conclude that the $D_N(\SU(p))$ theory can be mass-deformed to $(A_{N-1},A_{p-1})$. The relevant deformation corresponds to activating a mass for the $\SU(p)$ global symmetry associated with the full puncture. The IR fixed point of the corresponding RG flow is a class $\mathcal{S}$ theory of the same type, but on a sphere with only an irregular puncture; the regular puncture has been removed by the RG flow. Crucially, this process also modifies the irregular puncture relative to the one describing the UV $D_N(\SU(p))$ theory. 
							
							This modification of the puncture is consistent with the description in \cite{Giacomelli:2017ckh}, where it was argued that this flow is equivalent to the one discovered by Maruyoshi and Song \cite{Maruyoshi:2016tqk,Maruyoshi:2016aim,Agarwal:2016pjo}. However, while that construction involves breaking supersymmetry to $\mathcal{N}=1$, we find here that an RG flow with the same UV and IR fixed points can be achieved by activating a mass deformation for the global $\SU(p)$ symmetry. This operation manifestly preserves $\mathcal{N}=2$ supersymmetry throughout the entire RG flow. 
							
							We remark that, although the argument in this section implies the existence of this flow only for $N>p$, this restriction is not essential. The RG flow also exists for $N<p$. In that case, however, the FI deformation is more complex than simply deforming away the $T_{[1^p]}[\SU(p)]$ tail in the 3d mirror, as was done here.

							\section{Star-shaped Parent Quivers for \texorpdfstring{$D_p(\SU(N))$}{Dp(SU(N))}}
							\label{sec:StarShapedforDpSUN}
							
							We now present our main result: a general construction for the star-shaped parent quivers of $D_p(\SU(N))$ theories.
							
							\subsection{General Strategy: The Euclidean Algorithm} 
							\label{sec:generalstrategy}
							
							In this section, we propose a general form for the star-shaped quiver that deforms to the 3d mirror of any given $D_p(\SU(N))$ theory.
							
							As established in Section~\ref{sec:DeformingDpSUintoDNSU}, a $D_p(\SU(N))$ theory with $p \leq N$ can be mass-deformed into the $D_N(\SU(p))$ theory. This implies that both theories share the same star-shaped parent. It is therefore sufficient to construct the parent quiver for the case $p \leq N$.
							
							For convenience, we introduce the following shorthand notation for a full quiver tail of rank $r$:
							\begin{equation}\label{eq:fulltail}
								\begin{tikzpicture}[baseline=0,font=\footnotesize]
									\node[gauge,label=below:{$1$}] (u1B) at (0,0) {};
									\node[gauge,label=below:{$2$}] (u2B) [right=10mm of u1B] {};
									\node (dotsB) [right=10mm of u2B] {$\cdots$};
									\node[gauge,label=below:{$r$}] (uN1B) [right=10mm of dotsB]{};
									\draw[thick] (u1B) -- (u2B) -- (dotsB) -- (uN1B);
									\node [left=6mm of u1B] {$F_r:$};
								\end{tikzpicture}
							\end{equation}
							We propose the following ansatz for the parent star-shaped quiver of $D_p(\SU(N))$ with $p\leq N$:\footnote{In subsequent diagrams, an $F_1$ tail will often be depicted simply as \tikz[baseline=-0.5ex]{\node[gauge,label=above:{$1$}] (u1) {}; \node (u2) [right=2mm of u1] {}; \draw[thick] (u1)-- (u2);}.}
							\begin{equation} \label{ansatzDpSUN}
								\begin{tikzpicture}[baseline=35,font=\footnotesize]
									\node[gauge,label=below:{$N-x$}] (uBM) {};
									\node (dotsB1) [left=6mm of uBM] {$\cdots$};
									\node[gauge,label=below:{$2$}] (uB2) [left=6mm of dotsB1] {};
									\node[gauge,label=below:{$1$}] (uB1) [left=6mm of uB2] {};
									\node[gauge,label=below:{$N-x-(p-1)$}] (uN1) [right=24mm of uBM]{};
									\node (dotsA1) [right=12mm of uN1] {$\cdots$};
									\node[gauge,label=below:{$N-px$}] (uNp) [right=12mm of dotsA1]{};
									\node (U1d) [right=10mm of uNp] {$/\U(1)$};
									\node (u1) [above left=18mm of uBM] {$F_{\fh_1}$};
									\node(u12) [above right=18mm of uBM] {$F_{\fh_n}$};
									\node (dotsu1) [above=14mm of uBM] {$\cdots$};
									\draw[thick, decorate, decoration={brace,amplitude=5pt}]
									([yshift=4mm]u1.west) -- ([yshift=4mm]u12.east) node [black,pos=0.5,yshift=0.7cm] 
									{$\displaystyle\sum_{i=1}^n \fh_i = p$};
									\draw[thick] (uBM) -- (dotsB1) -- (uB2) -- (uB1);
									\draw[thick] (uBM) -- (uN1) -- (dotsA1) -- (uNp);
									\draw[thick] (uBM) -- (u1);
									\draw[thick] (uBM) -- (u12);
								\end{tikzpicture}
							\end{equation}
							where $x = \lfloor N/p \rfloor$. The quiver features a central node $\U(N-x)$ connected to several tails. The collection of full tails $F_{\fh_i}$ (where the largest node $\fh_i$ of each tail connects to the central node) is termed the \textit{bouquet}. The rightmost tail, which we denote by Tail $A'$ in \eqref{tailsA'B'}, has nodes whose ranks decrease by $p-1$ at each step, starting from $\U(N-x)$ down to $\U(N-px)$. The leftmost tail is denoted Tail $B'$ in \eqref{tailsA'B'}. This structure is analogous to the mirror theory of $D_p(\SU(N))$ itself.
							
							The next step in our strategy relies on the result discussed in Section~\ref{sec:DeformingDpSUintoDNSU}. We leverage this by considering a related theory, $D_N(\SU(p+kN))$ for $k \in \BZ_{\geq 1}$. Since $p+kN > N$, we can adapt our ansatz by applying the following substitutions to the parameters in \eqref{ansatzDpSUN}: 
							\begin{equation}
								p \rightarrow N\coma \quad N\rightarrow p+kN\coma \quad \text{and} \quad x \rightarrow \left\lfloor \frac{kN+p}{N} \right\rfloor  =k\fstop
							\end{equation}
							This yields the following ansatz for the parent star-shaped quiver of $D_N(\SU(p+kN))$: 
							\begin{equation} \label{ansatzDNSUp+kN}
								\begin{tikzpicture}[baseline=35,font=\footnotesize]
									\node[gauge,label=below:{$p+k(N-1)$}] (uBM) {};
									\node (dotsB1) [left=6mm of uBM] {$\cdots$};
									\node[gauge,label=below:{$2$}] (uB2) [left=6mm of dotsB1] {};
									\node[gauge,label=below:{$1$}] (uB1) [left=6mm of uB2] {};
									\node[gauge,label=below:{$p+(k-1)(N-1)$}] (uN1) [right=28mm of uBM]{};
									\node (dotsA1) [right=12mm of uN1] {$\cdots$};
									\node[gauge,label=below:{$p$}] (uNp) [right=12mm of dotsA1]{};
									\node (U1d) [right=10mm of uNp] {$/\U(1)$};
									\node (u1) [above left=18mm of uBM] {$F_{\fh_1}$};
									\node(u12) [above right=18mm of uBM] {$F_{\fh_{n'}}$};
									\node (dotsu1) [above=14mm of uBM] {$\cdots$};
									\draw[thick, decorate, decoration={brace,amplitude=5pt}]
									([yshift=4mm]u1.west) -- ([yshift=4mm]u12.east) node [black,pos=0.5,yshift=0.7cm] 
									{$\displaystyle\sum_{i=1}^{n'} \fh_i = N$};
									\draw[thick] (uBM) -- (dotsB1) -- (uB2) -- (uB1);
									\draw[thick] (uBM) -- (uN1) -- (dotsA1) -- (uNp);
									\draw[thick] (uBM) -- (u1);
									\draw[thick] (uBM) -- (u12);
								\end{tikzpicture}
							\end{equation}
							By assumption, $p \leq N$. We can therefore partition the bouquet of $n'$ tails, $\{F_{\fh_i}\}$, into two subsets such that
							\begin{equation}
								\sum_{i=1}^n \fh_i = p \coma \quad \sum_{i=n+1}^{n'}\fh_i = N-p\fstop
							\end{equation}
							We require the first subset, with $\sum_{i=1}^n \fh_i = p$, to be identical to the bouquet proposed for $D_p(\SU(N))$ in \eqref{ansatzDpSUN}. By requiring that the Higgs branch dimensions of the parent quivers \eqref{ansatzDpSUN} and \eqref{ansatzDNSUp+kN} match the ranks of the corresponding $D_p(\SU(N))$ and $D_N(\SU(p+kN))$ theories, we can uniquely determine the composition of the second subset. It must contain one copy of the tail $F_{N-px}$ and $(x-1)$ copies of the tail $F_p$. The complete star-shaped parent quiver for $D_N(\SU(p+kN))$ is therefore
							\begin{equation} \label{ansatzDNSUp+kNa}
								\begin{tikzpicture}[baseline=0,font=\footnotesize]
									\node[gauge,label={[xshift=1.5cm, yshift=-1.6mm]{$p+k(N-1)$}}] (uBM) {};
									\node (dotsB1) [left=6mm of uBM] {$\cdots$};
									\node[gauge,label=below:{$2$}] (uB2) [left=6mm of dotsB1] {};
									\node[gauge,label=below:{$1$}] (uB1) [left=6mm of uB2] {};
									\node[gauge,label=below:{$p+(k-1)(N-1)$}] (uN1) [right=28mm of uBM]{};
									\node (dotsA1) [right=12mm of uN1] {$\cdots$};
									\node[gauge,label=below:{$p$}] (uNp) [right=12mm of dotsA1]{};
									\node (U1d) [right=10mm of uNp] {$/\U(1)$};
									\node (u1) [above left=18mm of uBM] {$F_{\fh_1}$};
									\node(u12) [above right=18mm of uBM] {$F_{\fh_{n}}$};
									\node (u13) [below left=18mm of uBM] {$F_{p}$};
									\node(u14) [below right=18mm of uBM] {$F_{p}$};
									\node (dotsu1) [above=14mm of uBM] {$\cdots$};
									\node (dotsu2) [below=14mm of uBM] {$\cdots$};
									\node (FNpx) [right=6mm of u14] {$F_{N-px}$};
									\draw[thick, decorate, decoration={brace,amplitude=5pt}]
									([yshift=4mm]u1.west) -- ([yshift=4mm]u12.east) node [black,pos=0.5,yshift=0.7cm] 
									{$\displaystyle\sum_{i=1}^n \fh_i = p$};
									\draw[thick, decorate, decoration={brace,amplitude=5pt,mirror}]
									([yshift=-2mm]u13.west) -- ([yshift=-2mm]u14.east) node [black,pos=0.5,yshift=-0.7cm] 
									{$x-1$};
									\draw[thick] (uBM) -- (dotsB1) -- (uB2) -- (uB1);
									\draw[thick] (uBM) -- (uN1) -- (dotsA1) -- (uNp);
									\draw[thick] (uBM) -- (u1);
									\draw[thick] (uBM) -- (u12);
									\draw[thick] (uBM) -- (u13);
									\draw[thick] (uBM) -- (u14);
									\draw[thick] (uBM) -- (FNpx);
								\end{tikzpicture}
							\end{equation}
							This provides a direct method to construct the parent quiver for $D_N(\SU(p+kN))$ from the parent of $D_p(\SU(N))$. As a point of reference, we note that, for the special case of the Lagrangian theory $D_\mu(\SU(\alpha\mu))$, the mirror is a star-shaped quiver of the form \eqref{ansatzDpSUN} where the bouquet consists of $\mu$ copies of the $F_1$ tail, as shown in \cite[(5.26)]{Giacomelli:2020ryy}.
							
							To summarize, this procedure establishes a chain of deformations
							\bes{ \label{DpSUNtoDp'SUN'}
								D_p(\SU(N)) \,\, \rightarrow \,\,D_{p+kN}(\SU(N)) \,\, \rightarrow \,\,D_N (\SU(p+kN)) \equiv D_{p'}(\SU(N'))\fstop
							}
							where the second arrow relies on the result derived in Section~\ref{sec:DeformingDpSUintoDNSU}. The inverse of this process is
							\bes{
								D_{p'}(\SU(N')) \,\, \rightarrow \,\, D_{N' \, (\text{mod} \, p')} (\SU(p')) = D_p(\SU(N))~.
							}
							This observation motivates the use of the following sequence, generated via the {\it Euclidean algorithm}, which
							is a standard method to determine the greatest common divisor of two given integers ($N$ and $p$ in this case):
							\begin{equation} \label{euclideanalg}
								\begin{array}{rclrclrclcrcl}
									a_0 &= & N~,  & a_1  & = & p~,  & a_2  &=  & N \, (\text{mod} \, p)~,  & \ldots\,, & a_j  &=  & a_{j-2} \,\, (\text{mod}\,\, a_{j-1})~,  \\
									a_{\bar{n}}  &=  &\alpha \mu~,  & a_{\bar{n}+1}  &=  &\mu~,  &a_{\bar{n}+2}  &=  &0~, & \multicolumn{4}{l}{\text{ with }       \mu  =  \GCD(N,p)\,.}
								\end{array}
							\end{equation}
							This generates a strictly decreasing sequence of non-negative integers $(a_0, a_1, \ldots, a_{\bar{n}+1}, 0)$, which terminates at zero. The second-to-last term, $a_{\bar{n}+1}$, is the greatest common divisor, $\mu = \GCD(N,p)$. In terms of the theories, this algorithm defines a chain of deformations
							\bes{
								D_{a_1}(\SU(a_0)) \,\, \rightarrow \,\, D_{a_2}(\SU(a_1)) \,\, \rightarrow \,\, \cdots \,\, \rightarrow \,\, D_{\mu}(\SU(\alpha \mu))~.
							}
							Our main strategy for constructing the parent quiver of $D_p(\SU(N))$ is to reverse this chain:
							\bes{\label{seqdeform}
								\scalebox{0.95}{$
									D_{\mu}(\SU(\alpha \mu)) = D_{a_{\bar{n}+1}}(\SU(a_{\bar{n}})) \,\, \rightarrow \,\, D_{a_{\bar{n}}}(\SU(a_{\bar{n}-1})) \,\, \rightarrow \,\, \cdots\,\, \rightarrow \,\, D_{a_1}(\SU(a_0)) = D_p(\SU(N)) ~.
									$}
							}
							At each step of this reversed sequence, the parent quiver for a theory $D_{a_{q+1}}(\SU(a_q))$ takes the general form of \eqref{ansatzDpSUN}, with parameters $N=a_q$, $p=a_{q+1}$, and $x = \lfloor a_q / a_{q+1} \rfloor$. The composition of the bouquet is determined recursively:
							\begin{equation}\label{eq:recursionfulltails}
								\text{Bouquet for } D_{a_{q}}(\SU(a_{q-1})) = \text{Bouquet for } D_{a_{q+1}}(\SU(a_{q})) + F_{a_{q+2}} + \left[ (x-1)\times F_{a_{q+1}} \right]\coma
							\end{equation}
							where $(x-1) \times F_{a_{q+1}}$ denotes $(x-1)$ copies of the full tail $F_{a_{q+1}}$ attached to the central node. Using the fact that the bouquet for the base case, $D_\mu(\SU(\alpha\mu))$, consists of $\mu$ copies of the $F_1$ tail, we can solve this recursion relation to find the general form of the bouquet:
							\bes{
								\text{Bouquet for } D_{a_{q}}(\SU(a_{q-1})) &= \mu F_1 +\sum_{j=q}^{\bar{n}} \left[F_{a_{j+2}} + \left(\left \lfloor \frac{a_j}{a_{j+1}} \right \rfloor-1\right) F_{a_{j+1}}  \right] \\
								&= \mu F_1  -F_{a_{q+1}} + \sum_{j=q}^{\bar{n}} \left \lfloor \frac{a_j}{a_{j+1}} \right \rfloor F_{a_{j+1}}~.
							}
							For the specific theory $D_p(\SU(N))$, we have $a_0=N$ and $a_1=p$. Setting $q=1$ in the general solution yields:
							\bes{ \label{bouquetDpSUN}
								\text{Bouquet for } D_{p}(\SU(N)) &= \mu F_1 -F_{a_2}  +\sum_{j=1}^{\bar{n}} \left \lfloor \frac{a_j}{a_{j+1}} \right \rfloor F_{a_{j+1}} \\
								&= \mu F_1 +\left(\left \lfloor \frac{a_1}{a_2} \right \rfloor -1 \right ) F_{a_2} +\sum_{j=2}^{\bar{n}} \left \lfloor \frac{a_j}{a_{j+1}} \right \rfloor F_{a_{j+1}}\fstop
							}
							Here, $\mu = \GCD(N,p)$, whereas the sequence $a_j$ and the index $\bar{n}$ are determined by the Euclidean algorithm in \eqref{euclideanalg}. The expression in \eqref{bouquetDpSUN} fully specifies the bouquet content, which, when combined with the general structure of \eqref{ansatzDpSUN}, completely determines the star-shaped parent quiver for the $D_p(\SU(N))$ theory.
							
							This parent quiver is the 3d mirror of a class $\mathcal{S}$ theory of Type $A_{N-x-1}$ associated with a sphere with the following punctures \cite{Benini:2010uu}:
							\bes{
								\left[1^{N-x}\right], \quad \left[(p-1)^x,N-px\right]~, \quad c_i \times \left[N-x-\fh_i,1^{\fh_i}\right]~.
							}
							The parameters $c_i$ and $\fh_i$ are determined by writing the bouquet composition from \eqref{bouquetDpSUN} in the form $\sum_i c_i F_{\mathfrak{h}_i}$. The coefficient $c_i$ gives the multiplicity of the puncture type $[N-x-\fh_i, 1^{\fh_i}]$. This establishes our central claim: any $D_p(\SU(N))$ theory can be obtained from a sequence of mass deformations of the aforementioned class $\mathcal{S}$ theory. It is important to stress that this construction provides one possible star-shaped parent. We do not claim uniqueness, as it is known that some $D_p(\SU(N))$ theories can have multiple distinct parents.

							\subsubsection*{Example: \texorpdfstring{$D_{17 \mu}(\SU(26 \mu))$}{D17mu(SU(26mu))} with \texorpdfstring{$\mu \geq 1$}{mu>=1}}
							
							To illustrate this procedure, we determine a star-shaped parent quiver for the $D_{17 \mu}(\SU(26 \mu))$ theory. The Euclidean algorithm sequence from \eqref{euclideanalg} is
							\bes{
								(a_0, a_1, a_2, a_{\bar{n}=3}, a_4, a_5) = (26 \mu, 17 \mu, 9 \mu, 8 \mu, \mu, 0)\fstop
							}
							The corresponding sequence of theory deformations from \eqref{seqdeform} reads
							\bes{
								D_\mu(\SU(8\mu))\,\, \rightarrow \,\, D_{8\mu}(\SU(9\mu))\,\, \rightarrow \,\, D_{9\mu}(\SU(17\mu))\,\, \rightarrow \,\, D_{17\mu}(\SU(26\mu))\fstop
							}
							The bouquet composition of the star-shaped parent quiver at each step can be computed recursively using \eqref{eq:recursionfulltails}, which agrees with the general solution \eqref{bouquetDpSUN}. The results are summarized below:
							\bes{
								\begin{tabular}{c|c}
									Theory  & Bouquet  \\
									\hline
									$D_\mu (\SU(8\mu))$ & $\mu F_1$ \\ 
									$D_{8\mu} (\SU(9\mu))$ & $\mu F_1+7 F_\mu$ \\ 
									$D_{9\mu} (\SU(17\mu))$ & $\mu F_1+8 F_\mu$ \\ 
									$D_{17\mu} (\SU(26\mu))$ & $\mu F_1+8 F_\mu+F_{8\mu}$
								\end{tabular}
							}
							Therefore, $D_{17\mu}(\SU(26\mu))$ can be obtained by a series of mass deformations starting from the class $\CS$ theory of Type $A_{26\mu-2}=\SU(26\mu-1)$ associated with a sphere with the following punctures:
							\bes{
								[1^{26\mu-1}]~, \quad [17\mu-1,9\mu]~, \quad \mu \times [26\mu-2,1]~, \quad 8 \times [25\mu-1,1^\mu]~, \quad  [18\mu-1,1^{8\mu}]\fstop
							}
							
							\subsection{Motivation and Explanation for Ansatz \texorpdfstring{\eqref{ansatzDpSUN}}{Ansatz (4.2)}}
							\label{sec:AnsatzforDpSUN}
							
							This section provides the motivation and justification for the ansatz \eqref{ansatzDpSUN}, which was used to derive the star-shaped parent theory. A key validation of this ansatz is that the rank of the $D_p(\SU(N))$ theory (for $N\geq p$), given by
							\bes{
								\mathrm{rank}\,D_p(\SU(N)) = \frac{1}{2} \left[(N-1)(p-1) - (\mu-1) \right]~,\quad \mu= \GCD(N,p)\coma
							}
							is precisely equal to the dimension of the Higgs branch of the theory in \eqref{ansatzDpSUN} with the bouquet satisfying \eqref{bouquetDpSUN}. This equality provides a highly non-trivial check on the validity of our proposal. We now present further arguments to support it.
							
							The ansatz \eqref{ansatzDpSUN} consists of two main components:
							\begin{enumerate}
								\item The horizontal tails, shown as Tail $A'$ and Tail $B'$:
								\begin{equation} \label{tailsA'B'}
									\begin{tikzpicture}[baseline=15,font=\footnotesize]
										\node[gauge,label=below:{$N-px$}] (u1) {};
										\node (dots) [right=6mm of u1] {$\cdots$};
										\node[gauge,label=below:{$N-x-(p-1)$}] (uN1) [right=12mm of dots]{};
										\node[gauge,label=below:{$N-x$}] (uN2) [right=20mm of uN1]{};
										\draw[thick] (u1) -- (dots) -- (uN1) -- (uN2);
										\node [left=10mm of u1] {Tail $A'$:};
										\node[gauge,label=below:{$1$}] (u1B) [above=12mm of u1] {};
										\node[gauge,label=below:{$2$}] (u2B) [right=12mm of u1B] {};
										\node (dotsB) [right=6mm of u2B] {$\cdots$};
										\node[gauge,label=below:{$N-x$}] (uN1B) [above=12mm of uN2]{};
										\draw[thick] (u1B) -- (u2B) -- (dotsB) -- (uN1B);
										\node [left=10mm of u1B] {Tail $B'$:};
									\end{tikzpicture}
								\end{equation}
								\item The bouquet of tails, denoted by $F_{\fh_i}$.
							\end{enumerate}
							We will demonstrate that each component behaves as expected under specific deformations.
							
							First, we show that Tail $A'$ and Tail $B'$ in \eqref{tailsA'B'} can be deformed into Tail A and Tail B from \eqref{def:TailATailB}, respectively. This is achieved by activating FI parameters for the $\U(N-px)$ node in \eqref{ansatzDpSUN} and for the $\U(\fh_i)$ nodes of the $F_{\fh_i}$ tails, such that
							\begin{equation}
								\sum_j \fh_j = N-px\fstop
							\end{equation}
							This is always possible, since the sum of all $\fh_i$ is $p$, and $N-px$ is always less than $p$.\footnote{By definition $x=\left\lfloor \frac{N}{p}\right\rfloor$, so $x\leq \frac{N}{p}<x+1$. Multiplying by $p$, we obtain $xp\leq N<xp+p$, from which it follows that $0\leq N-px<p$. This is also clear by noting that $N-px$ is the element $a_2$ of the Euclidean sequence, which guarantees that either itself or its partitions are present in the bouquet of tails.} This deformation yields Tail A and Tail B connected to some of the $F_{\fh_i}$ tails. An explicit example of this process is the deformation from $D_{\mu}(\SU(\alpha \mu))$ to $D_{\mu}(\SU(\alpha \mu-1))$, discussed in Section~\ref{sec:DeformingDpSUNintoDpSUN-1}. If we view \eqref{eq:generalDmuSUalphamu} as a star-shaped parent quiver for $D_{\mu}(\SU(\alpha \mu-1))$ (with $N=\alpha \mu$, $p=\mu$, and $x=\alpha$), the required Tails A and B are explicitly present in \eqref{eq:generalDmuSUalphamu-2}. We also provide further examples in Sections~\ref{sec:ExampleDpSUmB1} and \ref{sec:DpSUNpN-2}.
							
							To motivate the second component of the ansatz, the bouquet of tails, we focus on the first transition in \eqref{DpSUNtoDp'SUN'}: from $D_p(\SU(N))$ to $D_{p+kN}(\SU(N))$. The remainder of this section argues that the star-shaped parent theory for the latter can be obtained by adding $k$ copies of the $F_{N-1}$ tail to the parent theory of the former. Since the second step of the transition was discussed in Section~\ref{sec:DeformingDpSUintoDNSU}, we only need to focus on this first step. Preliminary evidence for this claim comes from the rank difference:
							\bes{
								\mathrm{rank}\,D_{p+kN}(\SU(N))- \mathrm{rank}\,D_p(\SU(N)) = \frac{1}{2}kN(N-1)~.
							}
							This difference is precisely $k$ times the sum of the ranks of the gauge nodes in an $F_{N-1}$ tail.
							
							\subsubsection*{From \texorpdfstring{$D_{2n}(\SU(2))$}{D2n(SU(2))} to a Parent of \texorpdfstring{$D_{2n+2}(\SU(2))$}{D2n+2(SU(2))}}
							
							To provide further justification, let us examine the special case of $(A_1, D_{2n}) = D_{2n}(\SU(2))$. As noted in \cite[Section 4.3]{Xie:2012hs}, this theory can be embedded into the 4d $\mathcal{N}=2$ Lagrangian theory
							\bes{
								[1]-\underbrace{\SU(2) -\cdots-\SU(2)}_{n-1}-[2]\coma
							}
							which can be obtained via a mass deformation of the theory
							\bes{
								[2]-\underbrace{\SU(2) -\cdots-\SU(2)}_{n-1}-[2]\fstop
							}
							The latter is an $A_1$ class $\CS$ theory on a sphere with $n+2$ punctures. Its mirror theory is a star-shaped quiver with a central $\U(2)$ node and $(n+2)$ tails of type $F_1$.
							
							Now, let us consider the $D_{2n+2}(\SU(2))$ theory. To construct it, we couple the $T_2$ theory (the theory of four free hypermultiplets) to $D_{2n}(\SU(2))$ and gauge the diagonal $\SU(2)$ symmetry:
							\bes{
								\left([1]-\underbrace{\SU(2) -\cdots-\SU(2)}_{n-1}-[2]\,\,\, +\,\,\, [2]-[2]\right)///\SU(2)\fstop
							}
							This resulting theory can be obtained by a mass deformation of
							\bes{
								[2]-\underbrace{\SU(2) -\cdots-\SU(2)}_{n}-[2]\fstop
							}
							The mirror of this theory is a star-shaped quiver with $(n+3)$ tails of type $F_1$:
							\begin{equation}
								\begin{tikzpicture}[baseline=30,font=\footnotesize]
									\node[gauge,label=below:{$2$}] (uBM) {};
									\node[gauge,label=above:{$1$}] (u1) [above left=18mm of uBM] {};
									\node[gauge,label=above:{$1$}] (u12) [above right=18mm of uBM] {};
									\node (dotsu1) [above=14mm of uBM] {$\cdots$};
									\draw[thick, decorate,decoration={brace,amplitude=5pt}]
									([yshift=7mm]u1.west) -- ([yshift=7mm]u12.east) node [black,pos=0.5,yshift=0.4cm]
									{$n+3$};
									\draw[thick] (uBM) -- (u1);
									\draw[thick] (uBM) -- (u12);
									\node (U1d) [right=25mm of uBM] {$/\U(1)$};
								\end{tikzpicture}
							\end{equation}
							In summary, increasing $p$ by 2 in $D_{p=2n}(\SU(2))$ is equivalent to coupling the $T_2$ theory via $\SU(2)$ gauging. In the star-shaped parent theory, this corresponds to adding an $F_1$ tail.
							
							This concept generalizes to $D_p(\SU(N))$. We can increase $p$ by $N$ by coupling the $T_N$ theory to $D_p(\SU(N))$ via $\SU(N)$ gauging. In the star-shaped parent theory, this operation corresponds to adding an $F_{N-1}$ tail.
							
							To demonstrate this, let us construct the parent theory of $D_{k N}(\SU(N))$ starting from the mirror theory of $D_{N}(\SU(N))$. From \eqref{eq:generalDmuSUalphamu}, the mirror theory for $D_N(\SU(N))$ is
							\begin{equation} \label{mirrDNSUN}
								\begin{tikzpicture}[baseline=30,font=\footnotesize]
									\node[gauge,label=below:{$N-1$}] (uBM) {};
									\node (dotsB1) [left=6mm of uBM] {$\cdots$};
									\node[gauge,label=below:{$2$}] (uB2) [left=6mm of dotsB1] {};
									\node[gauge,label=below:{$1$}] (uB1) [left=6mm of uB2] {};
									\node (U1d) [right=25mm of uBM] {$/\U(1)$};
									\node[gauge,label=above:{$1$}] (u1) [above left=18mm of uBM] {};
									\node[gauge,label=above:{$1$}] (u12) [above right=18mm of uBM] {};
									\draw[thick, decorate,decoration={brace,amplitude=5pt}]
									([yshift=8mm]u1.west) -- ([yshift=8mm]u12.east) node [black,pos=0.5,yshift=0.4cm]
									{$N$};
									\draw[thick] (uBM) -- (dotsB1) -- (uB2) -- (uB1);
									\draw[thick] (uBM) -- (u1);
									\draw[thick] (uBM) -- (u12);
								\end{tikzpicture}
							\end{equation}
							Our goal is to find the star-shaped parent theory for $D_{kN}(\SU(N))$. To do this, we first consider a related theory:
							\begin{equation}
								\begin{tikzpicture}[baseline=30,font=\footnotesize]
									\node[gauge,label=below:{$N$}] (uBM) {};
									\node (dotsB1) [left=6mm of uBM] {$\cdots$};
									\node[gauge,label=below:{$2$}] (uB2) [left=6mm of dotsB1] {};
									\node[gauge,label=below:{$1$}] (uB1) [left=6mm of uB2] {};
									\node (U1d) [right=25mm of uBM] {$/\U(1)$};
									\node[gauge,label=above:{$1$}] (u1) [above left=18mm of uBM] {};
									\node[gauge,label=above:{$1$}] (u12) [above right=18mm of uBM] {};
									\draw[thick, decorate,decoration={brace,amplitude=5pt}]
									([yshift=8mm]u1.west) -- ([yshift=8mm]u12.east) node [black,pos=0.5,yshift=0.4cm]
									{$N$};
									\draw[thick] (uBM) -- (dotsB1) -- (uB2) -- (uB1);
									\draw[thick] (uBM) -- (u1);
									\draw[thick] (uBM) -- (u12);
								\end{tikzpicture}
							\end{equation}
							This theory is equivalent to \eqref{mirrDNSUN} modulo a set of free hypermultiplets, but its central $\U(N)$ node is now ``ugly" (i.e., it has fewer flavors than twice its rank). Now, we add $(k-1)$ copies of the $F_{N-1}$ tail to this quiver, which yields
							\begin{equation} \label{parentDkNSUN}
								\begin{tikzpicture}[baseline=0,font=\footnotesize]
									\node[gauge,label=right:{$N$}] (uBM) {};
									\node (dotsB1) [left=6mm of uBM] {$\cdots$};
									\node[gauge,label=below:{$2$}] (uB2) [left=6mm of dotsB1] {};
									\node[gauge,label=below:{$1$}] (uB1) [left=6mm of uB2] {};
									\node (U1d) [right=25mm of uBM] {$/\U(1)$};
									\node[gauge,label=above:{$1$}] (u1) [above left=18mm of uBM] {};
									\node[gauge,label=above:{$1$}] (u12) [above right=18mm of uBM] {};
									\node (u13) [below left=18mm of uBM] {$F_{N-1}$};
									\node(u14) [below right=18mm of uBM] {$F_{N-1}$};
									\node (dotsu1) [above=14mm of uBM] {$\cdots$};
									\node (dotsu2) [below=14mm of uBM] {$\cdots$};
									\draw[thick, decorate,decoration={brace,amplitude=5pt}]
									([yshift=7mm]u1.west) -- ([yshift=7mm]u12.east) node [black,pos=0.5,yshift=0.7cm]
									{$N$};
									\draw[thick, decorate,decoration={mirror,brace,amplitude=5pt}]
									([yshift=-2mm]u13.west) -- ([yshift=-2mm]u14.east) node [black,pos=0.5,yshift=-0.7cm]
									{$k-1$};
									\draw[thick] (uBM) -- (dotsB1) -- (uB2) -- (uB1);
									\draw[thick] (uBM) -- (u1);
									\draw[thick] (uBM) -- (u12);
									\draw[thick] (uBM) -- (u13);
									\draw[thick] (uBM) -- (u14);
								\end{tikzpicture}
							\end{equation}
							We propose that \eqref{parentDkNSUN} is the star-shaped parent theory for $D_{kN}(\SU(N))$. This claim will be substantiated in Sections~\ref{sec:D8SU4-example} and \ref{sec:D3kSU3-example}, where we explicitly demonstrate the deformations from \eqref{parentDkNSUN} to the known mirror theories of $D_8(\SU(4))$ and $D_{3k}(\SU(3))$ for generic $k$.
							
							\subsubsection*{General Case: From a Parent of \texorpdfstring{$D_p(\SU(N))$}{Dp(SU(N))} to a Parent of \texorpdfstring{$D_{p+N}(\SU(N))$}{Dp+N(SU(N))}}
							
							We can apply a similar procedure to the general ansatz \eqref{ansatzDpSUN}. First, we modify the quiver so that the central node becomes $\U(N)$ instead of $\U(N-x)$. This is achieved by transforming the left tail into an $F_{N-1} = 1-2-\cdots-(N-1)$ tail and increasing the rank of each node in the right tail from $N-x-j(p-1)$ to $N-jp$ (for $j=1, \ldots, x$). The resulting quiver is
							\begin{equation}
								\begin{tikzpicture}[baseline=30,font=\footnotesize]
									\node[gauge,label=below:{$N$}] (uBM) {};
									\node (dotsB1) [left=6mm of uBM] {$\cdots$};
									\node[gauge,label=below:{$2$}] (uB2) [left=6mm of dotsB1] {};
									\node[gauge,label=below:{$1$}] (uB1) [left=6mm of uB2] {};
									\node[gauge,label=below:{$N-p$}] (uN1) [right=24mm of uBM]{};
									\node (dotsA1) [right=12mm of uN1] {$\cdots$};
									\node[gauge,label=below:{$N-p x$}] (uNp) [right=12mm of dotsA1]{};
									\node (U1d) [right=10mm of uNp] {$/\U(1)$};
									\node (u1) [above left=18mm of uBM] {$F_{\fh_1}$};
									\node(u12) [above right=18mm of uBM] {$F_{\fh_n}$};
									\node (dotsu1) [above=14mm of uBM] {$\cdots$};
									\draw[thick, decorate,decoration={brace,amplitude=5pt}]
									([yshift=4mm]u1.west) -- ([yshift=4mm]u12.east) node [black,pos=0.5,yshift=0.7cm]
									{$\displaystyle\sum_{i=1}^n \fh_i = p$};
									\draw[thick] (uBM) -- (dotsB1) -- (uB2) -- (uB1);
									\draw[thick] (uBM) -- (uN1) -- (dotsA1) -- (uNp);
									\draw[thick] (uBM) -- (u1);
									\draw[thick] (uBM) -- (u12);
								\end{tikzpicture}
							\end{equation}
							This quiver is equivalent to the original ansatz \eqref{ansatzDpSUN} modulo free hypermultiplets, a consequence of its ``ugly'' central node. Adding an $F_{N-1}$ tail to this modified quiver results in the proposed star-shaped parent for $D_{p+N}(\SU(N))$:
							\begin{equation}
								\begin{tikzpicture}[baseline=0,font=\footnotesize]
									\node[gauge,label=below:{$N$}] (uBM) {};
									\node (dotsB1) [left=6mm of uBM] {$\cdots$};
									\node[gauge,label=below:{$2$}] (uB2) [left=6mm of dotsB1] {};
									\node[gauge,label=below:{$1$}] (uB1) [left=6mm of uB2] {};
									\node[gauge,label=below:{$N-p$}] (uN1) [right=24mm of uBM]{};
									\node (dotsA1) [right=12mm of uN1] {$\cdots$};
									\node[gauge,label=below:{$N-p x$}] (uNp) [right=12mm of dotsA1]{};
									\node (U1d) [right=10mm of uNp] {$/\U(1)$};
									\node (u1) [above left=18mm of uBM] {$F_{\fh_1}$};
									\node(u12) [above right=18mm of uBM] {$F_{\fh_n}$};
									\node(u13) [below left=18mm of uBM] {$F_{N-1}$};
									\node (dotsu1) [above=14mm of uBM] {$\cdots$};
									\draw[thick, decorate,decoration={brace,amplitude=5pt}]
									([yshift=4mm]u1.west) -- ([yshift=4mm]u12.east) node [black,pos=0.5,yshift=0.7cm]
									{$\displaystyle\sum_{i=1}^n \fh_i = p$};
									\draw[thick] (uBM) -- (dotsB1) -- (uB2) -- (uB1);
									\draw[thick] (uBM) -- (uN1) -- (dotsA1) -- (uNp);
									\draw[thick] (uBM) -- (u1);
									\draw[thick] (uBM) -- (u12);
									\draw[thick] (uBM) -- (u13);
								\end{tikzpicture}
							\end{equation}
							Crucially, this procedure preserves the overall structure of the ansatz \eqref{ansatzDpSUN}. The fact that adding an $F_{N-1}$ tail to the parent of $D_p(\SU(N))$ yields a theory of the same form for $D_{p+N}(\SU(N))$ provides strong motivation for the general validity of our ansatz.
							
							\section{Examples}
							\label{sec:examples}
							This section provides several explicit examples to demonstrate how the proposed star-shaped parent theories deform into the corresponding mirror theories of $D_p(\SU(N))$.
							
							\subsection{Deforming \texorpdfstring{\eqref{parentDkNSUN}$_{N=4, k=2}$}{} to the Mirror Theory for \texorpdfstring{$D_8(\SU(4))$}{D8SU4}}
							\label{sec:D8SU4-example}
							We begin by considering the case of $D_8(\SU(4))$, which corresponds to setting $N=4$ and $k=2$ in the parent quiver \eqref{parentDkNSUN}:
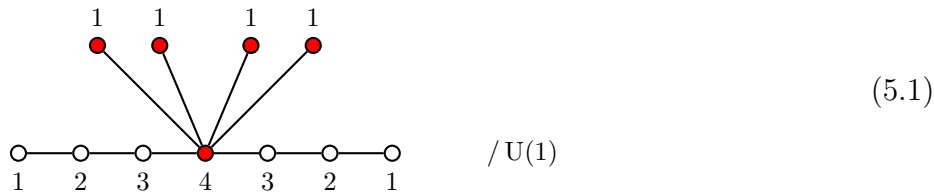
							\begin{equation}
								\begin{tikzpicture}[baseline=20,font=\footnotesize]
									\node[gauge,label=below:{$4$},fill=red] (uBM) {};
									\node[gauge,label=below:{$3$}] (dotsB1) [left=6mm of uBM] {};
									\node[gauge,label=below:{$2$}] (uB2) [left=6mm of dotsB1] {};
									\node[gauge,label=below:{$1$}] (uB1) [left=6mm of uB2] {};
									\node[gauge,label=below:{$3$}] (dotsA1) [right=6mm of uBM] {};
									\node[gauge,label=below:{$2$}] (uN1) [right=6mm of dotsA1]{};
									\node[gauge,label=below:{$1$}] (uNp) [right=6mm of uN1]{};
									\node (U1d) [right=10mm of uNp] {$/\U(1)$};
									\node[gauge,label=above:{$1$},fill=red] (u1) [above left=18mm of uBM] {};
									\node[gauge,label=above:{$1$},fill=red] (u2) [right=6mm of u1] {};
									\node[gauge,label=above:{$1$},fill=red] (u4) [above right=18mm of uBM] {};
									\node[gauge,label=above:{$1$},fill=red] (u3) [left=6mm of u4] {};
									\draw[thick] (uBM) -- (dotsB1) -- (uB2) -- (uB1);
									\draw[thick] (uBM) -- (dotsA1) -- (uN1) -- (uNp);
									\draw[thick] (uBM) -- (u1);
									\draw[thick] (uBM) -- (u2);
									\draw[thick] (uBM) -- (u3);
									\draw[thick] (uBM) -- (u4);
								\end{tikzpicture}
							\end{equation}
							where the nodes at which FI parameters are activated have been colored in {\color{red}{red}}. This deformation leads to the following quiver:
							
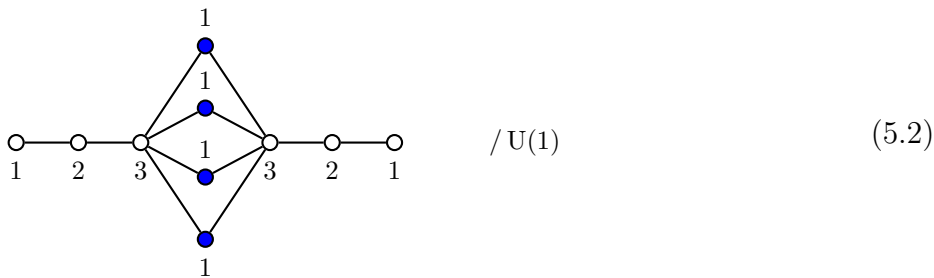
\begin{equation}\label{eq:D8SU4-intermidiate-1}
								\begin{tikzpicture}[baseline=0,font=\footnotesize]
									\node (uBM) {};
									\node[gauge,label=below:{$3$}] (dotsB1) [left=6mm of uBM] {};
									\node[gauge,label=below:{$2$}] (uB2) [left=6mm of dotsB1] {};
									\node[gauge,label=below:{$1$}] (uB1) [left=6mm of uB2] {};
									\node[gauge,label=below:{$3$}] (dotsA1) [right=6mm of uBM] {};
									\node[gauge,label=below:{$2$}] (uN1) [right=6mm of dotsA1]{};
									\node[gauge,label=below:{$1$}] (uNp) [right=6mm of uN1]{};
									\node (U1d) [right=10mm of uNp] {$/\U(1)$};
									\node[gauge,label=above:{$1$},fill=blue] (u1) [above=2mm of uBM] {};
									\node[gauge,label=above:{$1$},fill=blue] (u2) [above=6mm of u1] {};
									\node[gauge,label=above:{$1$},fill=blue] (u4) [below=2mm of uBM] {};
									\node[gauge,label=below:{$1$},fill=blue] (u3) [below=6mm of u4] {};
									\draw[thick]  (dotsB1) -- (uB2) -- (uB1);
									\draw[thick]  (dotsA1) -- (uN1) -- (uNp);
									\draw[thick] (dotsB1) -- (u1);
									\draw[thick] (dotsB1) -- (u2);
									\draw[thick] (dotsB1) -- (u3);
									\draw[thick] (dotsB1) -- (u4);
									\draw[thick] (dotsA1) -- (u1);
									\draw[thick] (dotsA1) -- (u2);
									\draw[thick] (dotsA1) -- (u3);
									\draw[thick] (dotsA1) -- (u4);
								\end{tikzpicture}
							\end{equation}
							where the {\color{blue}{blue}} nodes represent the balancing $\U(1)$ factors introduced by the deformation. Next, we activate FI parameters for the {\color{red}{red}} nodes shown in \eqref{eq:D8SU4-intermidiate-2}:
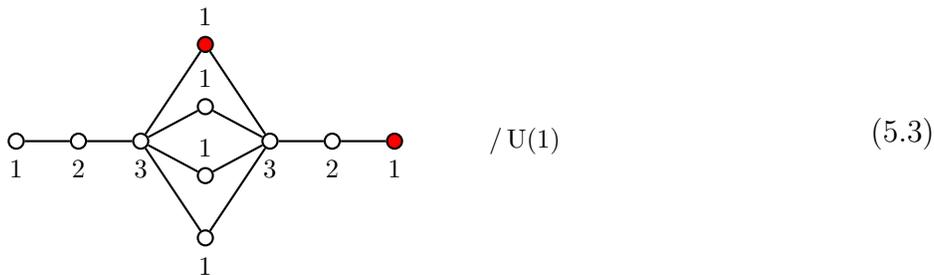
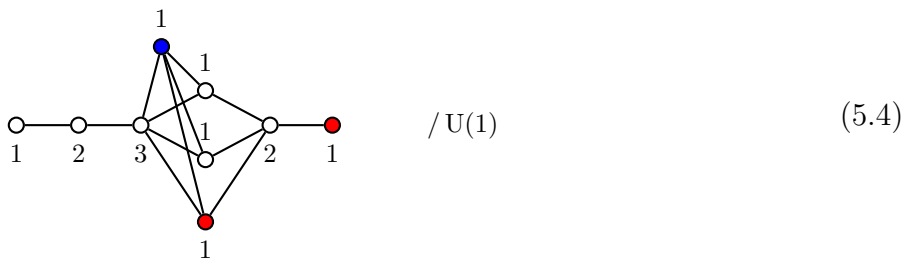
							\begin{equation}\label{eq:D8SU4-intermidiate-2}
								\begin{tikzpicture}[baseline=0,font=\footnotesize]
									\node (uBM) {};
									\node[gauge,label=below:{$3$}] (dotsB1) [left=6mm of uBM] {};
									\node[gauge,label=below:{$2$}] (uB2) [left=6mm of dotsB1] {};
									\node[gauge,label=below:{$1$}] (uB1) [left=6mm of uB2] {};
									\node[gauge,label=below:{$3$}] (dotsA1) [right=6mm of uBM] {};
									\node[gauge,label=below:{$2$}] (uN1) [right=6mm of dotsA1]{};
									\node[gauge,label=below:{$1$},fill=red] (uNp) [right=6mm of uN1]{};
									\node (U1d) [right=10mm of uNp] {$/\U(1)$};
									\node[gauge,label=above:{$1$}] (u1) [above=2mm of uBM] {};
									\node[gauge,label=above:{$1$},fill=red] (u2) [above=6mm of u1] {};
									\node[gauge,label=above:{$1$}] (u4) [below=2mm of uBM] {};
									\node[gauge,label=below:{$1$}] (u3) [below=6mm of u4] {};
									\draw[thick]  (dotsB1) -- (uB2) -- (uB1);
									\draw[thick]  (dotsA1) -- (uN1) -- (uNp);
									\draw[thick] (dotsB1) -- (u1);
									\draw[thick] (dotsB1) -- (u2);
									\draw[thick] (dotsB1) -- (u3);
									\draw[thick] (dotsB1) -- (u4);
									\draw[thick] (dotsA1) -- (u1);
									\draw[thick] (dotsA1) -- (u2);
									\draw[thick] (dotsA1) -- (u3);
									\draw[thick] (dotsA1) -- (u4);
								\end{tikzpicture}
							\end{equation}
							which yields
							\begin{equation}\label{eq:D8SU4-intermidiate-3}
								\begin{tikzpicture}[baseline=0,font=\footnotesize]
									\node (uBM) {};
									\node[gauge,label=below:{$3$}] (dotsB1) [left=6mm of uBM] {};
									\node[gauge,label=below:{$2$}] (uB2) [left=6mm of dotsB1] {};
									\node[gauge,label=below:{$1$}] (uB1) [left=6mm of uB2] {};
									\node[gauge,label=below:{$2$}] (dotsA1) [right=6mm of uBM] {};
									\node[gauge,label=below:{$1$},fill=red] (uN1) [right=6mm of dotsA1]{};
									\node (U1d) [right=10mm of uN1] {$/\U(1)$};
									\node[gauge,label=above:{$1$}] (u1) [above=2mm of uBM] {};
									\node[gauge,label=above:{$1$},fill=blue] (u2) [above left=6mm of u1] {};
									\node[gauge,label=above:{$1$}] (u4) [below=2mm of uBM] {};
									\node[gauge,label=below:{$1$},fill=red] (u3) [below=6mm of u4] {};
									\draw[thick]  (dotsB1) -- (uB2) -- (uB1);
									\draw[thick]  (dotsA1) -- (uN1);
									\draw[thick] (dotsB1) -- (u1);
									\draw[thick] (dotsB1) -- (u2);
									\draw[thick] (dotsB1) -- (u3);
									\draw[thick] (dotsB1) -- (u4);
									\draw[thick] (dotsA1) -- (u1);
									\draw[thick] (dotsA1) -- (u3);
									\draw[thick] (dotsA1) -- (u4);
									\draw[thick] (u2) -- (u1);
									\draw[thick] (u2) -- (u3);
									\draw[thick] (u2) -- (u4);
								\end{tikzpicture}
							\end{equation}
							where the {\color{red}{red}} nodes indicate the next set of FI parameters to be activated. The resulting quiver is
							\begin{equation}\label{eq:D8SU4-intermidiate-4}
								\begin{tikzpicture}[baseline=0,font=\footnotesize]
									\node (uBM) {};
									\node[gauge,label=below:{$3$}] (dotsB1) [left=8mm of uBM] {};
									\node[gauge,label=below:{$2$}] (uB2) [left=6mm of dotsB1] {};
									\node[gauge,label=below:{$1$}] (uB1) [left=6mm of uB2] {};
									\node[gauge,label=below:{$1$},fill=red] (dotsA1) [right=6mm of uBM] {};
									\node (U1d) [right=10mm of dotsA1] {$/\U(1)$};
									\node[gauge,label=above:{$1$},fill=red] (u1) [above=2mm of uBM] {};
									\node[gauge,label=above:{$1$}] (u2) [above left=6mm of u1] {};
									\node[gauge,label=above:{$1$}] (u4) [below=2mm of uBM] {};
									\node[gauge,label=below:{$1$},fill=blue] (u3) [below left =6mm of u4] {};
									\draw[thick]  (dotsB1) -- (uB2) -- (uB1);
									\draw[thick] (dotsB1) -- (u1);
									\draw[thick] (dotsB1) -- (u2);
									\draw[thick] (dotsB1) -- (u3);
									\draw[thick] (dotsB1) -- (u4);
									\draw[thick] (dotsA1) -- (u1);
									\draw[thick] (dotsA1) -- (u4);
									\draw[thick] (u2) -- (u1);
									\draw[thick] (u2) -- (u3);
									\draw[thick] (u2) -- (u4);
									\draw[thick] (u3) -- (u4);
									\draw[thick] (u3) -- (u1);
								\end{tikzpicture}
							\end{equation}
							A final deformation at the {\color{red}{red}} nodes yields the 3d mirror of $D_8(\SU(4))$:
							\begin{equation}\label{eq:D8SU4-mirror}
								\begin{tikzpicture}[baseline=0,font=\footnotesize]
									\node[gauge,label=below:{$3$}] (dotsB1) {};
									\node[gauge,label=below:{$2$}] (uB2) [left=6mm of dotsB1] {};
									\node[gauge,label=below:{$1$}] (uB1) [left=6mm of uB2] {};
									\node[gauge,label=above:{$1$}] (u1) [above right=6mm of dotsB1] {};
									\node[gauge,label=above:{$1$},fill=blue] (u2) [right=8mm of u1] {};
									\node[gauge,label=below:{$1$}] (u3) [below right=6mm of dotsB1] {};
									\node[gauge,label=below:{$1$}] (u4) [right =8mm of u3] {};
									\node (U1d) [right=25mm of dotsB1] {$/\U(1)$};
									\draw[thick]  (dotsB1) -- (uB2) -- (uB1);
									\draw[thick] (dotsB1) -- (u1);
									\draw[thick] (dotsB1) -- (u2);
									\draw[thick] (dotsB1) -- (u3);
									\draw[thick] (dotsB1) -- (u4);
									\draw[thick] (u2) -- (u1);
									\draw[thick] (u2) -- (u3);
									\draw[thick] (u2) -- (u4);
									\draw[thick] (u3) -- (u4);
									\draw[thick] (u3) -- (u1);
									\draw[thick] (u4) -- (u1);
								\end{tikzpicture}
							\end{equation}
							
							This example can also be used to explicitly demonstrate the result from Section~\ref{sec:DeformingANAp}. By activating FI parameters for the $\U(1)$ nodes of the full tail and one $\U(1)$ node in the complete graph, we can obtain the mirror of $(A_7,A_3)$. The initial step is
							\begin{equation}
								\begin{tikzpicture}[baseline=0,font=\footnotesize]
									\node[gauge,label=below:{$3$}] (dotsB1) {};
									\node[gauge,label=below:{$2$}] (uB2) [left=6mm of dotsB1] {};
									\node[gauge,label=below:{$1$},fill=red] (uB1) [left=6mm of uB2] {};
									\node[gauge,label=above:{$1$},fill=red] (u1) [above right=6mm of dotsB1] {};
									\node[gauge,label=above:{$1$}] (u2) [right=8mm of u1] {};
									\node[gauge,label=below:{$1$}] (u3) [below right=6mm of dotsB1] {};
									\node[gauge,label=below:{$1$}] (u4) [right =8mm of u3] {};
									\node (U1d) [right=25mm of dotsB1] {$/\U(1)$};
									\draw[thick]  (dotsB1) -- (uB2) -- (uB1);
									\draw[thick] (dotsB1) -- (u1);
									\draw[thick] (dotsB1) -- (u2);
									\draw[thick] (dotsB1) -- (u3);
									\draw[thick] (dotsB1) -- (u4);
									\draw[thick] (u2) -- (u1);
									\draw[thick] (u2) -- (u3);
									\draw[thick] (u2) -- (u4);
									\draw[thick] (u3) -- (u4);
									\draw[thick] (u3) -- (u1);
									\draw[thick] (u4) -- (u1);
								\end{tikzpicture}
							\end{equation}
							which yields
							\begin{equation}
								\begin{tikzpicture}[baseline=0,font=\footnotesize]
									\node[gauge,label=below:{$2$}] (dotsB1) {};
									\node[gauge,label=below:{$1$}] (uB2) [left=6mm of dotsB1] {};
									\node[gauge,label=above:{$1$},fill=blue] (u1) [above right=10mm of dotsB1] {};
									\node[gauge,label=above:{$1$}] (u2) [right=12mm of u1] {};
									\node[gauge,label=below:{$1$}] (u3) [below right=10mm of dotsB1] {};
									\node[gauge,label=below:{$1$}] (u4) [right =12mm of u3] {};
									\node (U1d) [right=30mm of dotsB1] {$/\U(1)$};
									\draw[thick]  (dotsB1) -- (uB2);
									\draw[thick] (dotsB1) -- (u2);
									\draw[thick] (dotsB1) -- (u3);
									\draw[thick] (dotsB1) -- (u4);
									\draw[thick] (u2) -- node[above,midway,sloped] {$2$} (u1);
									\draw[thick] (u2) --  (u3);
									\draw[thick] (u2) --  (u4);
									\draw[thick] (u3) -- (u4);
									\draw[thick] (u3) -- node[above,midway,sloped] {$2$} (u1);
									\draw[thick] (u4) -- node[above,pos=0.3,sloped] {$2$} (u1);
								\end{tikzpicture}
							\end{equation}
							We note that the balancing node is no longer connected to the full tail. Repeating this deformation with another $\U(1)$ node from the complete graph gives
							\begin{equation}
								\begin{tikzpicture}[baseline=0,font=\footnotesize]
									\node[gauge,label=below:{$1$}] (dotsB1) {};
									\node[gauge,label=above:{$1$}] (u1) [above right=10mm of dotsB1] {};
									\node[gauge,label=above:{$1$},fill=blue] (u2) [right=12mm of u1] {};
									\node[gauge,label=below:{$1$}] (u3) [below right=10mm of dotsB1] {};
									\node[gauge,label=below:{$1$}] (u4) [right =12mm of u3] {};
									\node (U1d) [right=30mm of dotsB1] {$/\U(1)$};
									\draw[thick] (dotsB1) -- (u3);
									\draw[thick] (dotsB1) -- (u4);
									\draw[thick] (u2) -- node[above,midway,sloped] {$2$} (u1);
									\draw[thick] (u2) -- node[above,pos=0.3,sloped] {$2$} (u3);
									\draw[thick] (u2) -- node[above,midway,sloped] {$2$} (u4);
									\draw[thick] (u3) -- (u4);
									\draw[thick] (u3) -- node[above,midway,sloped] {$2$} (u1);
									\draw[thick] (u4) -- node[above,pos=0.3,sloped] {$2$} (u1);
								\end{tikzpicture}
							\end{equation}
							A final deformation results in the complete graph associated with the mirror of $(A_7,A_3)$, consistent with the findings in \cite[(4.11)]{Giacomelli:2020ryy}:\footnote{In \cite[(4.11)]{Giacomelli:2020ryy}, the authors write the mirror obtained by closing the full puncture of the $D_p(\SU(N))$ theory, while here we are performing a mass deformation of the theory as described in Section \ref{sec:DeformingANAp}. In order to match the mirror, one needs to consider that $(A_3,A_7)$, which is equivalent to $(A_7,A_3)$, in \cite{Giacomelli:2020ryy} is obtained from $D_{12}(\SU(8))$.}
							\begin{equation}

							\end{equation}

							\subsection{Deforming \texorpdfstring{\eqref{parentDkNSUN}$_{N=3}$}{} to the Mirror Theory for \texorpdfstring{$D_{3k}(\SU(3))$}{D3kSU3}}
							\label{sec:D3kSU3-example}
							
							We now consider the case of $D_{3k}(\SU(3))$, which corresponds to setting $N=3$ in the parent quiver \eqref{parentDkNSUN} for a generic integer $k$:
							\begin{equation}
								%
							\end{equation}
							Activating FI parameters at the {\color{red}{red}} nodes and rebalancing the quiver with three {\color{blue}{blue}} $\U(1)$ nodes yields
							\begin{equation}
								%
							\end{equation}
							Next, we activate FI parameters for one of the {\color{blue}{blue}} $\U(1)$ nodes and one of the $\U(1)$ nodes in an $F_2$ tail, leading to
							\begin{equation}
								%
							\end{equation}
							The FI deformation at the {\color{red}{red}} nodes increases the multiplicity of the edge connecting the {\color{blue}{blue}} and {\color{red}{red}} nodes by one, resulting in
							\begin{equation}
								%
							\end{equation}
							By activating FI parameters for the top-right $\U(1)$ node and another $\U(1)$ from an $F_2$ tail, we can reach a quiver where the three top $\U(1)$ nodes are connected by edges of multiplicity two:
							\begin{equation}
								%
							\end{equation}
							By reiterating this sequence of deformations, we ultimately arrive at
							\begin{equation}
								%
							\end{equation}
							with $\alpha + \beta + \gamma = k-1$. At this stage, activating FI parameters for the {\color{red}{red}} nodes increases the multiplicity of the edge labeled $\beta + \gamma$ by one and reduces the number of nodes in the set labeled $\alpha$ by one. The resulting quiver is
							\begin{equation}
								%
							\end{equation}
							Similarly, an FI deformation at the next set of {\color{red}{red}} nodes increases the multiplicity of the edge labeled $\alpha + \gamma$ by one, while reducing the number of nodes in the set labeled $\beta$ by one:
							\begin{equation}
								%
							\end{equation}
							Activating FI parameters at the {\color{red}{red}} nodes increases the multiplicity of the edge labeled $\alpha + \beta$ by one and decreases the number of nodes in the set labeled $\gamma$ by one:
							\begin{equation}
								%
							\end{equation}
							Reiterating this sequence of FI deformations ultimately yields the 3d mirror of $D_{3 k}(\SU(3))$:
							\begin{equation} \label{3dmirrD3kSU3}
								%
							\end{equation}
							Finally, we note that $D_{3 k}(\SU(3))$ can be further deformed to $D_{3 k-1}(\SU(3))$. This is achieved by activating FI parameters for two $\U(1)$ nodes in the complete graph of \eqref{3dmirrD3kSU3}, which gives
							\begin{equation} 
								%
							\end{equation}
							Implementing this deformation on the two {\color{red}{red}} $\U(1)$ nodes yields the 3d mirror of $D_{3k-1}(\SU(3))$:
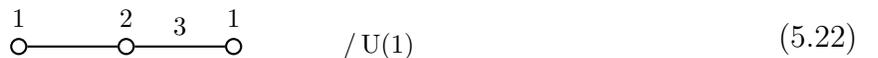
							\begin{equation} 
								%
							\end{equation}
							Further deforming the full tail with the single $\U(1)$ on the right leads to the 3d mirror of $(A_{3k-2},A_2)$, which consists solely of $3k-2$ free hypermultiplets (see, e.g., \cite[Table 2 or (6.35)-(6.36)]{Carta:2021whq}). 
							
							\subsection{Star-shaped Parent of \texorpdfstring{$D_p(\SU(N))$}{} with \texorpdfstring{$p\leq N$}{} and \texorpdfstring{$m_B=1$}{}}
							\label{sec:ExampleDpSUmB1}
							
							We now analyze the case where $m_B = (N- p x)/\mu=1$, with $x= \lfloor N/p \rfloor \geq 1$ and $\mu=\GCD(N,p)$. This implies $m_A = p/\mu-1$. The Euclidean algorithm sequence \eqref{euclideanalg} is
							\bes{
								a_0 = N =\mu+ px~, \quad a_1 =p~, \quad a_2 = \mu~, \quad a_3 = 0\fstop
							}
							From \eqref{bouquetDpSUN}, the bouquet consists of $\mu F_1 + (p/\mu-1) F_\mu = \mu F_1 + m_A F_\mu$. Noting that $N-x =M+1$, the ansatz \eqref{ansatzDpSUN} becomes
							\begin{equation}
								\begin{tikzpicture}[baseline=0,font=\footnotesize]
									\node[gauge,label={[xshift=0.9cm, yshift=-1.6mm]{$M+1$}}] (uBM) {};
									\node (dotsB1) [left=6mm of uBM] {$\cdots$};
									\node[gauge,label=below:{$2$}] (uB2) [left=6mm of dotsB1] {};
									\node[gauge,label=below:{$1$}] (uB1) [left=6mm of uB2] {};
									\node[gauge,label=below:{$M+1-(p-1)$}] (uN1) [right=28mm of uBM] {};
									\node (dotsA1) [right=12mm of uN1] {$\cdots$};
									\node[gauge,label=below:{$\mu$},fill=red] (uNp) [right=12mm of dotsA1] {};
									\node (U1d) [right=10mm of uNp] {$/\U(1)$};
									\node[gauge,label=above:{$1$},fill=red] (u1) [above left=18mm of uBM] {};
									\node[gauge,label=above:{$1$},fill=red] (u12) [above right=18mm of uBM] {};
									\node (u13) [below left=18mm of uBM] {$F_{\mu}$};
									\node(u14) [below right=18mm of uBM] {$F_{\mu}$};
									\node (dotsu1) [above=14mm of uBM] {$\cdots$};
									\node (dotsu2) [below=14mm of uBM] {$\cdots$};
									\draw[thick] [decorate,decoration={brace,amplitude=5pt},xshift=0cm, yshift=0cm]
									([yshift=7mm]u1.west) -- ([yshift=7mm]u12.east) node [black,pos=0.5,yshift=0.4cm] 
									{$\mu$};
									\draw[thick] [decorate,decoration={mirror,brace,amplitude=5pt},xshift=0cm, yshift=0cm]
									([yshift=-2mm]u13.west) -- ([yshift=-2mm]u14.east) node [black,pos=0.5,yshift=-0.7cm] 
									{$m_A$};
									\draw[thick] (uBM) -- (dotsB1) -- (uB2) -- (uB1);
									\draw[thick] (uBM) -- (uN1) -- (dotsA1) -- (uNp);
									\draw[thick] (uBM) -- (u1);
									\draw[thick] (uBM) -- (u12);
									\draw[thick] (uBM) -- (u13);
									\draw[thick] (uBM) -- (u14);
								\end{tikzpicture}
							\end{equation}
							where the {\color{red}{red}} nodes indicate where FI parameters are activated. We will now show that this quiver deforms into the mirror theory for the corresponding $D_p(\SU(N))$, adopting the notation from Appendix~\ref{sec:DpSUNmirrorsp<=N}.
							
							The deformation yields the following quiver:
							\begin{equation}\label{eq:DpSU-intermidiatemb=1}
								\begin{tikzpicture}[baseline=40,font=\footnotesize]
									\node[gauge,label=below:{$p-1$}] (u1) {};
									\node[gauge,label=below:{$2(p-1)$}] (u2) [right=12mm of u1] {};
									\node (dots) [right=6mm of u2] {$\cdots$};
									\node[gauge,label=below:{$(x-1)(p-1)$}] (uN1) [right=6mm of dots]{};
									\node[gauge,label=below:{$x(p-1)$}] (uN2) [right=18mm of uN1]{};
									\draw[thick] (u1) -- (u2) -- (dots) -- (uN1) -- (uN2);
									\node[gauge,label=below:{$1$}] (u1B) [above=12mm of u1] {};
									\node[gauge,label=below:{$2$}] (u2B) [right=12mm of u1B] {};
									\node (dotsB) [right=6mm of u2B] {$\cdots$};
									\node[gauge,label=below:{$M-1$}] (uN1B) [right=6mm of dotsB]{};
									\node[gauge,label=below left:{$M$}] (uN2B) [right=18mm of uN1B]{};
									\draw[thick] (u1B) -- (u2B) -- (dotsB) -- (uN1B) -- (uN2B);
									\node[gauge,label=above:{$1$},fill=blue] (u1) [above left=18mm of uN2B] {};
									\node[gauge,label=above:{$1$},fill=blue] (u12) [above right=18mm of uN2B] {};
									\node (dotsu1) [above=14mm of uN2B] {$\cdots$};
									\draw[thick] (u1) -- (uN2B);
									\draw[thick] (u12) -- (uN2B);
									\node[gauge,label=below:{$\mu$}] (u13) [right =30mm of uN2] {};
									\node (dotsmu1) [right =6mm of u13] {$\cdots$};
									\node[gauge,label=below:{$1$}] (u14) [right =6mm of dotsmu1] {};
									\draw[thick] (u13) -- (dotsmu1) -- (u14);
									\node[gauge,label=below:{$\mu$}] (u15) [above =12mm of u13] {};
									\node (dotsmu2) [right =6mm of u15] {$\cdots$};
									\node[gauge,label=below:{$1$}] (u16) [right =6mm of dotsmu2] {};
									\draw[thick] (u15) -- (dotsmu2) -- (u16);
									\node (vdots) [above=2mm of dotsmu1] {$\vdots$};
									\draw[thick][decorate,decoration={brace,amplitude=5pt},xshift=0cm, yshift=0cm] ([yshift=7mm]u1.west) -- ([yshift=7mm]u12.east) node [black,pos=0.5,yshift=0.4cm]  {$\mu$};
									\draw[thick][decorate,decoration={brace,amplitude=5pt},xshift=0cm, yshift=0cm] ([xshift=4mm]u16.north) -- ([xshift=4mm]u14.south) node [black,pos=0.5,xshift=0.7cm]  {$m_A$};
									\draw[thick] (uN2) -- (uN2B);
									\draw[thick] (u1) -- (u13);
									\draw[thick] (u1) -- (u15);
									\draw[thick] (u12) -- (u13);
									\draw[thick] (u12) -- (u15);
									\draw[thick] (uN2) -- (u13);
									\draw[thick] (uN2) -- (u15);
									\node (U1d) [right=15mm of u14] {$/\U(1)$};
								\end{tikzpicture}
							\end{equation}
							We note that the two tails on the left of \eqref{eq:DpSU-intermidiatemb=1} are already Tail A and Tail B as defined in Appendix~\ref{sec:DpSUNmirrorsp<=N}. Tail A is connected to the $m_A$ tails of type $F_\mu$, while Tail B is only connected to the bouquet of $\U(1)$ nodes that arise from rebalancing the quiver. All $\U(1)$s in the bouquet are also connected to the $\U(\mu)$ nodes of the $F_\mu$ tails. To obtain the mirror of $D_p(\SU(N))$, we must remove the $m_A$ tails of type $F_\mu$. This process will result in a complete graph of $\U(1)$ nodes connected to Tail A with multiplicity $m_A$ and to Tail B with multiplicity one. Let us demonstrate this deformation by activating the following FI parameters:
							\begin{equation}\label{eq:DpSU-intermidiatemb=1-2}

							\end{equation}
							After the deformation, the resulting quiver is
							\begin{equation}\label{eq:DpSU-intermidiatemb=1-3}
								%
							\end{equation}
							The balancing node in {\color{blue}{blue}} is now connected to Tail A, but not to the remaining $F_{\mu-1}$ tails. We can proceed by eliminating the $F_{\mu-1}$ tail by activating FI parameters at another node in the bouquet of $\mu-1$ $\U(1)$s. At the end of this process, the bouquet of $\mu$ $\U(1)$ nodes will be connected to Tail A with multiplicity one:
							\begin{equation}\label{eq:DpSU-intermidiatemb=1-4}
								%
							\end{equation}
							We note that the bouquet of $\U(1)$s is now also connected by an internal edge. This procedure is iterated until all $F_\mu$ tails are removed. Each iteration increases the multiplicity of the edges connecting the $\U(1)$ bouquet to Tail A by one, as well as the multiplicity of the edges within the complete graph. The final resulting quiver is
							\begin{equation}
								%
							\end{equation}
							This is precisely the quiver \eqref{eq:generalDpSU-p<=N} for the case $m_B=1$. 
							
							\subsection{Star-shaped Parent of \texorpdfstring{$D_p(\SU(N))$}{} with \texorpdfstring{$p\leq N$}{} and \texorpdfstring{$m_A=1$}{}}
							\label{sec:ExampleDpSUmA1}
							
							We now consider the case where $m_A = \left[p(1+x)-N\right]/\mu=1$, with $x= \lfloor N/p \rfloor \geq 1$ and $\mu=\GCD(N,p)$. Since $m_A+m_B=p/\mu$, we have $m_B = (N-px)/\mu = p/\mu-1$. The Euclidean algorithm sequence \eqref{euclideanalg} depends on whether $m_B$ is also equal to $1$: 
							\bes{
								\begin{array}{rclrcll}
									m_A&= &1\,, & m_B& \neq &1\,: &a_0 = N = p(1+x)-\mu~, \quad a_1 =p~, \quad a_2 = p-\mu = \alpha \mu\,\,\, (\alpha \in \BZ_{\geq 2})~,\\
									& & & & & &a_3 = \mu~, \quad a_4=0~,\\
									m_A& =& 1\,, & m_B& =& 1\,: &a_0 = N =\mu+ px~, \quad a_1 =p~, \quad a_2 = \mu~, \quad a_3 = 0~.
								\end{array}
							}
							The case $m_A=m_B=1$ was discussed in the preceding section. Let us focus on the case $m_A=1$ and $m_B\neq 1$. From the definitions, this implies $p=(\alpha+1)\mu$ and $m_B = \alpha$. Using \eqref{bouquetDpSUN}, we find that the bouquet consists of $\mu F_1 + (\lfloor(\alpha+1)/\alpha \rfloor-1) F_{\alpha \mu} + \alpha F_\mu = \mu F_1 + \alpha F_\mu = \mu F_1 + m_B F_\mu$. Writing $N-x =M+1$, the ansatz \eqref{ansatzDpSUN} becomes:
							\begin{equation}
								%
							\end{equation}
							where we use again the notation from Appendix~\ref{sec:DpSUNmirrorsp<=N}. This quiver is similar to the one in Section~\ref{sec:ExampleDpSUmB1}, but the previous deformation strategy cannot be used here, because there are not enough $\U(1)$ nodes in the bouquet to reduce the right tail. However, there are precisely $m_B$ tails of type $F_\mu$, whose $\U(\mu)$ nodes can be used to activate FI parameters along with the $\U(\mu m_B)$ node. After the deformation, the resulting quiver is
							\begin{equation}\label{eq:DpSU-intermidiatema=1}
								%
							\end{equation}
							This quiver is analogous to \eqref{eq:DpSU-intermidiatemb=1}, but with the $m_B$ tails of type $F_\mu$ connected to Tail B instead of Tail A. From this point, one can proceed as before by forming a complete graph of the $\U(1)$ nodes and attaching it to Tails A and B by eliminating the $m_B$ tails of type $F_\mu$. Each elimination of an $F_\mu$ tail increases the multiplicity of the edges connecting Tail B to the complete graph of $\U(1)$s, and also the multiplicity of the edges within the complete graph itself. Finally, the resulting quiver is
							\begin{equation}
								%
							\end{equation}
							This is precisely the quiver \eqref{eq:generalDpSU-p<=N} for $m_A=1$. 
							
							\subsection{Star-shaped Parent of \texorpdfstring{$D_p(\SU(N))$}{} with \texorpdfstring{$p = N-2$}{}}
							\label{sec:DpSUNpN-2}
							
							In this section, we explicitly show that the star-shaped parent quiver for $D_{N-2}(\SU(N))$, as prescribed in Section~\ref{sec:StarShapedforDpSUN}, deforms into the correct 3d mirror theory for $N \ge 4$. The relevant parameters are:
							\bes{ \label{paramNm2}
								\scalebox{0.9}{$
									\begin{array}{rlrlrl}
										\mu =& \begin{cases}
											2 & \text{$N$ even} \\
											1 & \text{$N$ odd}
										\end{cases}\coma \,
										&x=\left\lfloor \frac{N}{N-2}\right\rfloor =& \begin{cases}
											2 & N = 4 \\
											1 & N > 4
										\end{cases}\coma \,
										&M =& \begin{cases}
											1 & N = 4 \\
											N-2 & N > 4
										\end{cases}\coma \\ & & & & & \\
										m_A =& \begin{cases}
											1 & N=4 \\
											\frac{N}{2}-2 & \text{$N > 4$ even} \\ 
											N - 4 & \text{$N$ odd}
										\end{cases}\coma \,
										&m_B =& \begin{cases}
											0 & N=4 \\
											1 & \text{$N > 4$ even} \\
											2 & \text{$N$ odd}
										\end{cases}\coma \,
										& H_\text{\tiny free} =& \begin{cases}
											0 & \text{$N$ even} \\
											\frac{N-5}{2} & \text{$N$ odd}
										\end{cases}\fstop
									\end{array}
									$}
							}
							Observe that, for $N=4$ and for even $N > 4$, we have $m_A = 1$ and $m_B = 1$, respectively. Therefore, these cases are included in the analysis described in Sections~\ref{sec:ExampleDpSUmB1} and~\ref{sec:ExampleDpSUmA1}. We can thus focus just on the cases with odd $N$.
							
							For odd $N$, the integer sequence \eqref{euclideanalg} reads
							\bes{
								(a_0, a_1, a_{\bar{n}=2}, a_{\bar{n}+1=3}, a_{\bar{n}+2=4}) = (N, N-2, 2, 1, 0)~,
							}
							which corresponds to the sequence of deformations
							\bes{
								D_1(\SU(2))\,\, \rightarrow \,\, D_{2}(\SU(N-2))\,\, \rightarrow \,\, D_{N-2}(\SU(N))\fstop
							}
							Using \eqref{bouquetDpSUN}, the bouquet is given by
							\bes{ \label{bouqueNm2}
								\text{Bouquet for $D_{N-2}(\SU(N))$ } = 3 F_1 +\left( \frac{N-5}{2} \right)  F_{2}\fstop
							}
							Therefore, from \eqref{paramNm2} and \eqref{bouqueNm2}, the star-shaped parent quiver for $D_{N-2}(\SU(N))$ with odd $N$ is given by
							
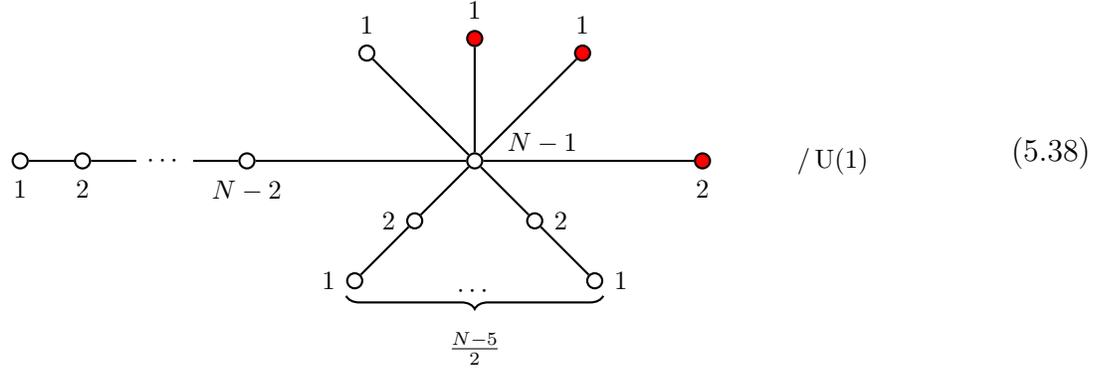
\begin{equation}
								\begin{tikzpicture}[baseline=0,font=\footnotesize]
									\node[gauge,label={[xshift=0.9cm, yshift=-1.6mm]{$N-1$}}] (uBM) {};
									\node[gauge,label=below:{$N-2$}] (uBMl) [left=28mm of uBM] {};
									\node (dotsB1) [left=6mm of uBMl] {$\cdots$};
									\node[gauge,label=below:{$2$}] (uB2) [left=6mm of dotsB1] {};
									\node[gauge,label=below:{$1$}] (uB1) [left=6mm of uB2] {};
									\node[gauge,label=below:{$2$},fill=red] (uN1) [right=28mm of uBM] {};
									\node (U1d) [right=10mm of uN1] {$/\U(1)$};
									\node[gauge,label=above:{$1$}] (u1) [above left=18mm of uBM] {};
									\node[gauge,label=above:{$1$},fill=red] (u12) [above right=18mm of uBM] {};
									\node[gauge,label=above:{$1$},fill=red] (u13) [above=14mm of uBM] {};
									\node[gauge,label=left:{$2$}] (u21) [below left=9mm of uBM] {};
									\node[gauge,label=right:{$2$}] (u22) [below right=9mm of uBM] {};
									\node[gauge,label=left:{$1$}] (u14) [below left=9mm of u21] {};
									\node[gauge,label=right:{$1$}](u15) [below right=9mm of u22] {};
									\node (dotsu2) [below=14mm of uBM] {$\cdots$};
									\draw[thick] [decorate,decoration={mirror,brace,amplitude=5pt},xshift=0cm, yshift=0cm] ([yshift=-2mm]u14.west) -- ([yshift=-2mm]u15.east) node [black,pos=0.5,yshift=-0.7cm] {$\frac{N-5}{2}$};
									\draw[thick] (uBM) --(uBMl) -- (dotsB1) -- (uB2) -- (uB1);
									\draw[thick] (uBM) -- (uN1);
									\draw[thick] (uBM) -- (u1);
									\draw[thick] (uBM) -- (u12);
									\draw[thick] (uBM) -- (u13);
									\draw[thick] (uBM) -- (u21) -- (u14);
									\draw[thick] (uBM) -- (u22) -- (u15);
								\end{tikzpicture}
							\end{equation}
							Activating FI parameters on the {\color{red}{red}} nodes yields the deformed quiver
							
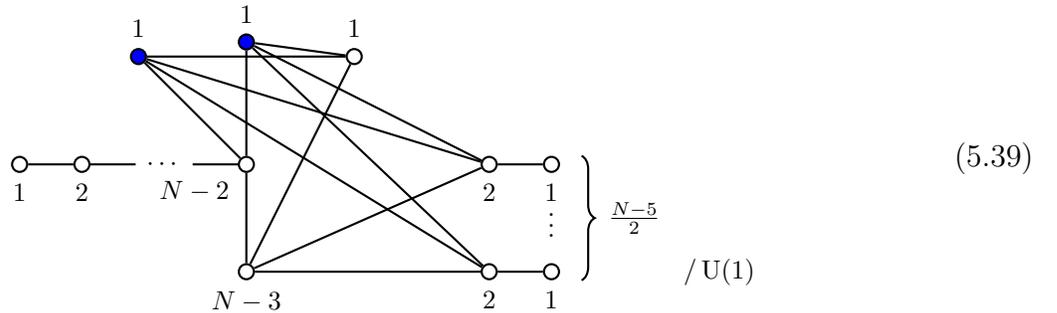
\begin{equation}
								\begin{tikzpicture}[baseline=40,font=\footnotesize]
									\node[gauge,label=below:{$N-3$}] (uN3) {};
									\node[gauge,label=below left:{$N-2$}] (uN) [above=12mm of uN3]{};
									\node (dotsB) [left=6mm of uN] {$\cdots$};
									\node[gauge,label=below:{$2$}] (u2B) [left=6mm of dotsB] {};
									\node[gauge,label=below:{$1$}] (u1B) [left=6mm of u2B] {};
									\draw[thick] (u1B) -- (u2B) -- (dotsB) -- (uN);
									\node[gauge,label=above:{$1$},fill=blue] (u1) [above left=18mm of uN] {};
									\node[gauge,label=above:{$1$}] (u12) [above right=18mm of uN] {};
									\node[gauge,label=above:{$1$},fill=blue] (u13) [above=14mm of uN] {};
									\draw[thick] (u1) -- (uN);
									\draw[thick] (u13) -- (uN);
									\draw[thick] (u12) -- (uN3);
									\draw[thick] (u1) -- (u12);
									\draw[thick] (u13) -- (u12);
									\node[gauge,label=below:{$2$}] (u14) [right =30mm of uN3] {};
									\node[gauge,label=below:{$1$}] (u15) [right =6mm of u14] {};
									\draw[thick] (u14) -- (u15);
									\node[gauge,label=below:{$2$}] (u16) [above =12mm of u14] {};
									\node[gauge,label=below:{$1$}] (u17) [right =6mm of u16] {};
									\draw[thick] (u16) -- (u17);
									\node (vdots) [above=2mm of u15] {$\vdots$};
									\draw[thick][decorate,decoration={brace,amplitude=5pt},xshift=0cm, yshift=0cm] ([xshift=4mm]u17.north) -- ([xshift=4mm]u15.south) node [black,pos=0.5,xshift=0.7cm]  {$\frac{N-5}{2}$};
									\draw[thick] (uN) -- (uN3);
									\draw[thick] (u1) -- (u14);
									\draw[thick] (u13) -- (u14);
									\draw[thick] (u1) -- (u16);
									\draw[thick] (u13) -- (u16);
									\draw[thick] (uN3) -- (u14);
									\draw[thick] (uN3) -- (u16);
									\node (U1d) [right=15mm of u15] {$/\U(1)$};
								\end{tikzpicture}
							\end{equation}
							To obtain a quiver analogous to those in Sections~\ref{sec:ExampleDpSUmB1} and~\ref{sec:ExampleDpSUmA1}, we first reduce the number of $\U(1)$s in the bouquet to $\mu=1$. We activate FI parameters at the following {\color{red}{red}} nodes:
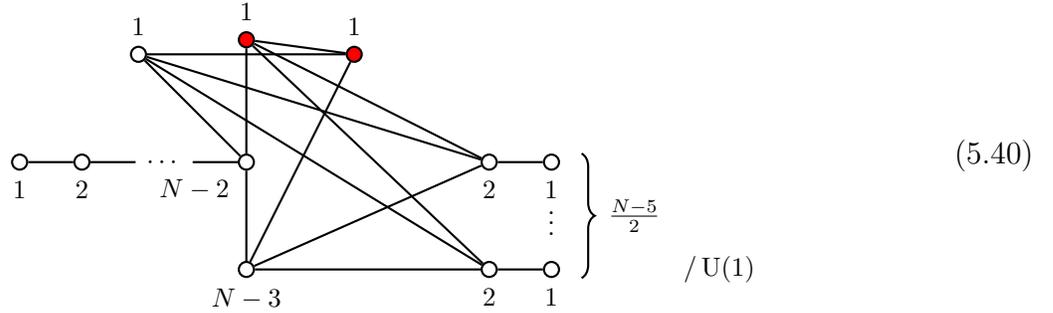
							\begin{equation}
								\begin{tikzpicture}[baseline=40,font=\footnotesize]
									\node[gauge,label=below:{$N-3$}] (uN3) {};
									\node[gauge,label=below left:{$N-2$}] (uN) [above=12mm of uN3]{};
									\node (dotsB) [left=6mm of uN] {$\cdots$};
									\node[gauge,label=below:{$2$}] (u2B) [left=6mm of dotsB] {};
									\node[gauge,label=below:{$1$}] (u1B) [left=6mm of u2B] {};
									\draw[thick] (u1B) -- (u2B) -- (dotsB) -- (uN);
									\node[gauge,label=above:{$1$}] (u1) [above left=18mm of uN] {};
									\node[gauge,label=above:{$1$},fill=red] (u12) [above right=18mm of uN] {};
									\node[gauge,label=above:{$1$},fill=red] (u13) [above=14mm of uN] {};
									\draw[thick] (u1) -- (uN);
									\draw[thick] (u13) -- (uN);
									\draw[thick] (u12) -- (uN3);
									\draw[thick] (u1) -- (u12);
									\draw[thick] (u13) -- (u12);
									\node[gauge,label=below:{$2$}] (u14) [right =30mm of uN3] {};
									\node[gauge,label=below:{$1$}] (u15) [right =6mm of u14] {};
									\draw[thick] (u14) -- (u15);
									\node[gauge,label=below:{$2$}] (u16) [above =12mm of u14] {};
									\node[gauge,label=below:{$1$}] (u17) [right =6mm of u16] {};
									\draw[thick] (u16) -- (u17);
									\node (vdots) [above=2mm of u15] {$\vdots$};
									\draw[thick][decorate,decoration={brace,amplitude=5pt},xshift=0cm, yshift=0cm] ([xshift=4mm]u17.north) -- ([xshift=4mm]u15.south) node [black,pos=0.5,xshift=0.7cm]  {$\frac{N-5}{2}$};
									\draw[thick] (uN) -- (uN3);
									\draw[thick] (u1) -- (u14);
									\draw[thick] (u13) -- (u14);
									\draw[thick] (u1) -- (u16);
									\draw[thick] (u13) -- (u16);
									\draw[thick] (uN3) -- (u14);
									\draw[thick] (uN3) -- (u16);
									\node (U1d) [right=15mm of u15] {$/\U(1)$};
								\end{tikzpicture}
							\end{equation}
							resulting in
							
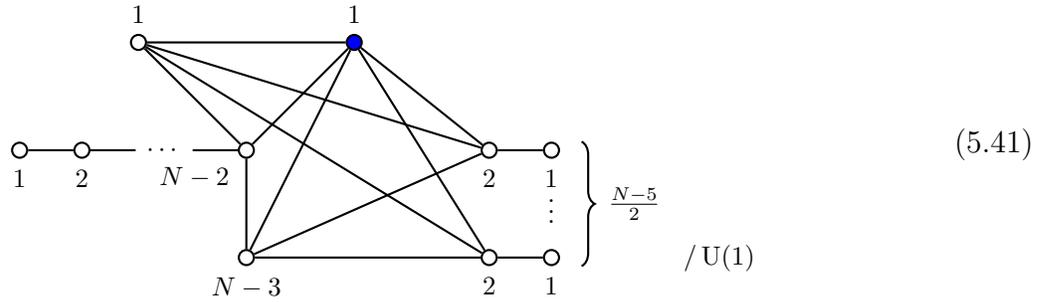
\begin{equation}
								\begin{tikzpicture}[baseline=40,font=\footnotesize]
									\node[gauge,label=below:{$N-3$}] (uN3) {};
									\node[gauge,label=below left:{$N-2$}] (uN) [above=12mm of uN3]{};
									\node (dotsB) [left=6mm of uN] {$\cdots$};
									\node[gauge,label=below:{$2$}] (u2B) [left=6mm of dotsB] {};
									\node[gauge,label=below:{$1$}] (u1B) [left=6mm of u2B] {};
									\draw[thick] (u1B) -- (u2B) -- (dotsB) -- (uN);
									\node[gauge,label=above:{$1$},fill=blue] (u12) [above right=18mm of uN] {};
									\node[gauge,label=below:{$2$}] (u14) [right =30mm of uN3] {};
									\node[gauge,label=below:{$1$}] (u15) [right =6mm of u14] {};
									\draw[thick] (u14) -- (u15);
									\node[gauge,label=below:{$2$}] (u16) [above =12mm of u14] {};
									\node[gauge,label=below:{$1$}] (u17) [right =6mm of u16] {};
									\draw[thick] (u16) -- (u17);
									\node (vdots) [above=2mm of u15] {$\vdots$};
									\node[gauge,label=above:{$1$}] (u1) [above left=18mm of uN] {};
									\draw[thick][decorate,decoration={brace,amplitude=5pt},xshift=0cm, yshift=0cm] ([xshift=4mm]u17.north) -- ([xshift=4mm]u15.south) node [black,pos=0.5,xshift=0.7cm]  {$\frac{N-5}{2}$};
									\draw[thick] (uN) -- (uN3);
									\draw[thick] (uN3) -- (u14);
									\draw[thick] (uN3) -- (u16);
									\draw[thick] (u12) -- (uN);
									\draw[thick] (u12) -- (uN3);
									\draw[thick] (u12) -- (u16);
									\draw[thick] (u12) -- (u14);
									\draw[thick] (u1) -- (uN);
									\draw[thick] (u1) -- (u14);
									\draw[thick] (u1) -- (u16);
									\draw[thick] (u1) -- (u12);
									\node (U1d) [right=15mm of u15] {$/\U(1)$};
								\end{tikzpicture}
							\end{equation}
							Repeating the deformation at the two remaining $\U(1)$ nodes in the bouquet leads to
							
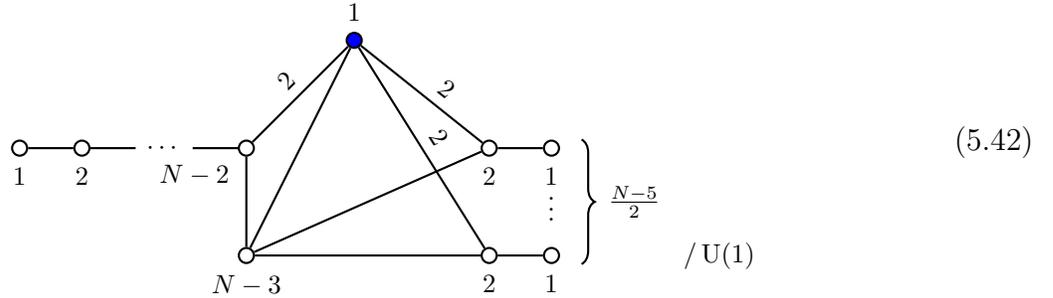
\begin{equation}
								\begin{tikzpicture}[baseline=40,font=\footnotesize]
									\node[gauge,label=below:{$N-3$}] (uN3) {};
									\node[gauge,label=below left:{$N-2$}] (uN) [above=12mm of uN3]{};
									\node (dotsB) [left=6mm of uN] {$\cdots$};
									\node[gauge,label=below:{$2$}] (u2B) [left=6mm of dotsB] {};
									\node[gauge,label=below:{$1$}] (u1B) [left=6mm of u2B] {};
									\draw[thick] (u1B) -- (u2B) -- (dotsB) -- (uN);
									\node[gauge,label=above:{$1$},fill=blue] (u12) [above right=18mm of uN] {};
									\node[gauge,label=below:{$2$}] (u14) [right =30mm of uN3] {};
									\node[gauge,label=below:{$1$}] (u15) [right =6mm of u14] {};
									\draw[thick] (u14) -- (u15);
									\node[gauge,label=below:{$2$}] (u16) [above =12mm of u14] {};
									\node[gauge,label=below:{$1$}] (u17) [right =6mm of u16] {};
									\draw[thick] (u16) -- (u17);
									\node (vdots) [above=2mm of u15] {$\vdots$};
									\draw[thick][decorate,decoration={brace,amplitude=5pt},xshift=0cm, yshift=0cm] ([xshift=4mm]u17.north) -- ([xshift=4mm]u15.south) node [black,pos=0.5,xshift=0.7cm]  {$\frac{N-5}{2}$};
									\draw[thick] (uN) -- (uN3);
									\draw[thick] (uN3) -- (u14);
									\draw[thick] (uN3) -- (u16);
									\draw[thick] (u12) -- node[above,sloped,midway] {$2$} (uN);
									\draw[thick] (u12) -- (uN3);
									\draw[thick] (u12) -- node[above,sloped,pos=0.6] {$2$} (u16);
									\draw[thick] (u12) -- node[above,sloped,midway] {$2$} (u14);
									\node (U1d) [right=15mm of u15] {$/\U(1)$};
									\node (Hfree) [below=23mm of dotsB] {$+1$ free hypermultiplet.};
								\end{tikzpicture}
							\end{equation}
							We now remove the $(N-5)/2$ tails of type $F_2$ by activating FI parameters at the single $\U(1)$ in the bouquet and the $\U(1)$ in each $F_2$ tail, \ie
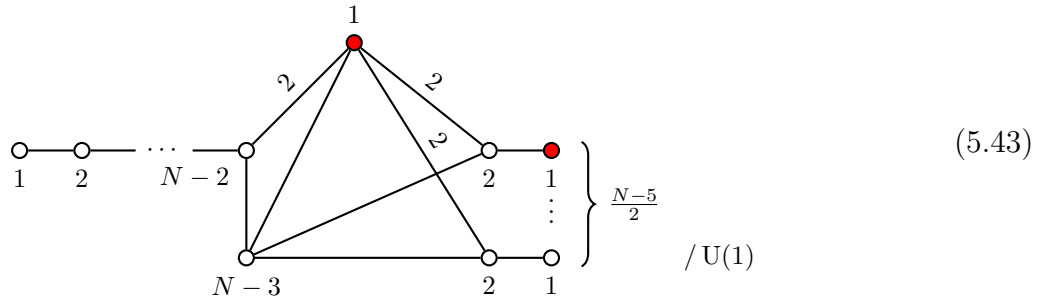
							\begin{equation}
								\begin{tikzpicture}[baseline=40,font=\footnotesize]
									\node[gauge,label=below:{$N-3$}] (uN3) {};
									\node[gauge,label=below left:{$N-2$}] (uN) [above=12mm of uN3]{};
									\node (dotsB) [left=6mm of uN] {$\cdots$};
									\node[gauge,label=below:{$2$}] (u2B) [left=6mm of dotsB] {};
									\node[gauge,label=below:{$1$}] (u1B) [left=6mm of u2B] {};
									\draw[thick] (u1B) -- (u2B) -- (dotsB) -- (uN);
									\node[gauge,label=above:{$1$},fill=red] (u12) [above right=18mm of uN] {};
									\node[gauge,label=below:{$2$}] (u14) [right =30mm of uN3] {};
									\node[gauge,label=below:{$1$}] (u15) [right =6mm of u14] {};
									\draw[thick] (u14) -- (u15);
									\node[gauge,label=below:{$2$}] (u16) [above =12mm of u14] {};
									\node[gauge,label=below:{$1$},fill=red] (u17) [right =6mm of u16] {};
									\draw[thick] (u16) -- (u17);
									\node (vdots) [above=2mm of u15] {$\vdots$};
									\draw[thick][decorate,decoration={brace,amplitude=5pt},xshift=0cm, yshift=0cm] ([xshift=4mm]u17.north) -- ([xshift=4mm]u15.south) node [black,pos=0.5,xshift=0.7cm]  {$\frac{N-5}{2}$};
									\draw[thick] (uN) -- (uN3);
									\draw[thick] (uN3) -- (u14);
									\draw[thick] (uN3) -- (u16);
									\draw[thick] (u12) -- node[above,sloped,midway] {$2$} (uN);
									\draw[thick] (u12) -- (uN3);
									\draw[thick] (u12) -- node[above,sloped,midway] {$2$} (u16);
									\draw[thick] (u12) -- node[above,sloped,midway] {$2$} (u14);
									\node (U1d) [right=15mm of u15] {$/\U(1)$};
								\end{tikzpicture}
							\end{equation}
							Such deformation leads to the removal of a single $F_2$ tail as follows:
							\begin{equation}
								\begin{tikzpicture}[baseline=40,font=\footnotesize]
									\node[gauge,label=below:{$N-3$}] (uN3) {};
									\node[gauge,label=below left:{$N-2$}] (uN) [above=12mm of uN3]{};
									\node (dotsB) [left=6mm of uN] {$\cdots$};
									\node[gauge,label=below:{$2$}] (u2B) [left=6mm of dotsB] {};
									\node[gauge,label=below:{$1$}] (u1B) [left=6mm of u2B] {};
									\draw[thick] (u1B) -- (u2B) -- (dotsB) -- (uN);
									\node[gauge,label=above:{$1$},fill=blue] (u12) [above right=18mm of uN] {};
									\node[gauge,label=above:{$1$}] (u13) [right=12mm of u12] {};
									\node[gauge,label=below:{$2$}] (u14) [right =30mm of uN3] {};
									\node[gauge,label=below:{$1$}] (u15) [right =6mm of u14] {};
									\draw[thick] (u14) -- (u15);
									\node[gauge,label=below:{$2$}] (u16) [above =12mm of u14] {};
									\node[gauge,label=below:{$1$}] (u17) [right =6mm of u16] {};
									\draw[thick] (u16) -- (u17);
									\node (vdots) [above=2mm of u15] {$\vdots$};
									\draw[thick][decorate,decoration={brace,amplitude=5pt},xshift=0cm, yshift=0cm] ([xshift=4mm]u17.north) -- ([xshift=4mm]u15.south) node [black,pos=0.5,xshift=1cm]  {$\frac{N-5}{2}-1$};
									\draw[thick] (uN) -- (uN3);
									\draw[thick] (uN3) -- (u14);
									\draw[thick] (uN3) -- (u16);
									\draw[thick] (u12) -- node[above,sloped,midway] {$2$} (uN);
									\draw[thick] (u12) -- node[above,sloped,midway] {$2$} (uN3);
									\draw[thick] (u12) -- node[above,sloped,pos=0.6] {$2$} (u16);
									\draw[thick] (u12) -- node[above,sloped,midway] {$2$} (u14);
									\draw[thick] (u13) -- (u12);
									\draw[thick] (u13) -- (uN3);
									\node (U1d) [right=15mm of u15] {$/\U(1)$};
									\node (Hfree) [below=23mm of dotsB] {$+1$ free hypermultiplet.};
								\end{tikzpicture}
							\end{equation}
							Activating FI parameters at the two $\U(1)$ nodes above, namely
							\begin{equation}
								\begin{tikzpicture}[baseline=40,font=\footnotesize]
									\node[gauge,label=below:{$N-3$}] (uN3) {};
									\node[gauge,label=below left:{$N-2$}] (uN) [above=12mm of uN3]{};
									\node (dotsB) [left=6mm of uN] {$\cdots$};
									\node[gauge,label=below:{$2$}] (u2B) [left=6mm of dotsB] {};
									\node[gauge,label=below:{$1$}] (u1B) [left=6mm of u2B] {};
									\draw[thick] (u1B) -- (u2B) -- (dotsB) -- (uN);
									\node[gauge,label=above:{$1$},fill=red] (u12) [above right=18mm of uN] {};
									\node[gauge,label=above:{$1$},fill=red] (u13) [right=12mm of u12] {};
									\node[gauge,label=below:{$2$}] (u14) [right =30mm of uN3] {};
									\node[gauge,label=below:{$1$}] (u15) [right =6mm of u14] {};
									\draw[thick] (u14) -- (u15);
									\node[gauge,label=below:{$2$}] (u16) [above =12mm of u14] {};
									\node[gauge,label=below:{$1$}] (u17) [right =6mm of u16] {};
									\draw[thick] (u16) -- (u17);
									\node (vdots) [above=2mm of u15] {$\vdots$};
									\draw[thick][decorate,decoration={brace,amplitude=5pt},xshift=0cm, yshift=0cm] ([xshift=4mm]u17.north) -- ([xshift=4mm]u15.south) node [black,pos=0.5,xshift=1cm]  {$\frac{N-5}{2}-1$};
									\draw[thick] (uN) -- (uN3);
									\draw[thick] (uN3) -- (u14);
									\draw[thick] (uN3) -- (u16);
									\draw[thick] (u12) -- node[above,sloped,midway] {$2$} (uN);
									\draw[thick] (u12) -- node[above,sloped,midway] {$2$} (uN3);
									\draw[thick] (u12) -- node[above,sloped,pos=0.6] {$2$} (u16);
									\draw[thick] (u12) -- node[above,sloped,midway] {$2$} (u14);
									\draw[thick] (u13) -- (u12);
									\draw[thick] (u13) -- (uN3);
									\node (U1d) [right=15mm of u15] {$/\U(1)$};
									\node (Hfree) [below=23mm of dotsB] {$+1$ free hypermultiplet,};
								\end{tikzpicture}
							\end{equation}
							results in the following quiver:
							\begin{equation}
								\begin{tikzpicture}[baseline=40,font=\footnotesize]
									\node[gauge,label=below:{$N-3$}] (uN3) {};
									\node[gauge,label=below left:{$N-2$}] (uN) [above=12mm of uN3]{};
									\node (dotsB) [left=6mm of uN] {$\cdots$};
									\node[gauge,label=below:{$2$}] (u2B) [left=6mm of dotsB] {};
									\node[gauge,label=below:{$1$}] (u1B) [left=6mm of u2B] {};
									\draw[thick] (u1B) -- (u2B) -- (dotsB) -- (uN);
									\node[gauge,label=above:{$1$},fill=blue] (u12) [above right=18mm of uN] {};
									\node[gauge,label=below:{$2$}] (u14) [right =30mm of uN3] {};
									\node[gauge,label=below:{$1$}] (u15) [right =6mm of u14] {};
									\draw[thick] (u14) -- (u15);
									\node[gauge,label=below:{$2$}] (u16) [above =12mm of u14] {};
									\node[gauge,label=below:{$1$}] (u17) [right =6mm of u16] {};
									\draw[thick] (u16) -- (u17);
									\node (vdots) [above=2mm of u15] {$\vdots$};
									\draw[thick][decorate,decoration={brace,amplitude=5pt},xshift=0cm, yshift=0cm] ([xshift=4mm]u17.north) -- ([xshift=4mm]u15.south) node [black,pos=0.5,xshift=1cm]  {$\frac{N-5}{2}-1$};
									\draw[thick] (uN) -- (uN3);
									\draw[thick] (uN3) -- (u14);
									\draw[thick] (uN3) -- (u16);
									\draw[thick] (u12) -- node[above,sloped,midway] {$2$} (uN);
									\draw[thick] (u12) -- node[above,sloped,midway] {$3$} (uN3);
									\draw[thick] (u12) -- node[above,sloped,pos=0.6] {$2$} (u16);
									\draw[thick] (u12) -- node[above,sloped,midway] {$2$} (u14);
									\node (U1d) [right=15mm of u15] {$/\U(1)$};
									\node (Hfree) [below=23mm of dotsB] {$+1$ free hypermultiplet.};
								\end{tikzpicture}
							\end{equation}
							We conclude that removing an $F_2$ tail increases the multiplicity of the edge connecting the $\U(1)$ bouquet to the $\U(N-3)$ node by two and generates one free hypermultiplet. This process can be repeated $(N-5)/2$ times, eventually leading to:
							\begin{equation}
								\begin{tikzpicture}[baseline=0,font=\footnotesize]
									\node[gauge,label=below:{$N-3$}] (uN3) {};
									\node[gauge,label=below:{$N-2$}] (uN) [left=24mm of uN3]{};
									\node (dotsB) [left=6mm of uN] {$\cdots$};
									\node[gauge,label=below:{$2$}] (u2B) [left=6mm of dotsB] {};
									\node[gauge,label=below:{$1$}] (u1B) [left=6mm of u2B] {};
									\draw[thick] (u1B) -- (u2B) -- (dotsB) -- (uN);
									\node[gauge,label=above:{$1$}] (u12) [above right=18mm of uN] {};
									\draw[thick] (uN) -- (uN3);
									\draw[thick] (u12) -- node[above,sloped,midway] {$2$} (uN);
									\draw[thick] (u12) -- node[above right,midway] {$N-4$} (uN3);
									\node (U1d) [right=30mm of uN3] {$/\U(1)$};
									\node (Hfree) [below=8mm of dotsB] {$+\frac{N-5}{2}$ free hypermultiplets.};
								\end{tikzpicture}
							\end{equation}
							This is precisely the mirror of $D_{N-2}(\SU(N))$ for odd $N$. We can also use this example to explicitly test the result from Section~\ref{sec:DeformingDpSUintoDNSU}. By activating FI parameters at the two $\U(N-3)$ nodes, we obtain
							\begin{equation}
								\begin{tikzpicture}[baseline=0,font=\footnotesize]
									\node[gauge,label=below:{$1$},fill=red] (uN3) {};
									\node[gauge,label=below:{$1$},fill=red] (u12) [left=24mm of uN3]{};
									\node[gauge,label=below:{$N-3$},fill=blue] (uN) [left=24mm of u12]{};
									\node (dotsB) [left=6mm of uN] {$\cdots$};
									\node[gauge,label=below:{$2$}] (u2B) [left=6mm of dotsB] {};
									\node[gauge,label=below:{$1$}] (u1B) [left=6mm of u2B] {};
									\draw[thick] (u1B) -- (u2B) -- (dotsB) -- (uN);
									\draw[thick] (u12) -- node[above,sloped,midway] {$N-2$} (uN);
									\draw[thick] (u12) -- node[above, sloped,midway] {$2$} (uN3);
									\node (U1d) [right=30mm of uN3] {$/\U(1)$};
									\node (Hfree) [below=8mm of dotsB] {$+\frac{N-5}{2}$ free hypermultiplets,};
								\end{tikzpicture}
							\end{equation}
							where the $\U(N-3)$ node in {\color{blue}{blue}} arises from rebalancing after the deformation, and we have colored in {\color{red}{red}} the two $\U(1)$ nodes for the next deformation. The final quiver is:
							\begin{equation}
								\begin{tikzpicture}[baseline=0,font=\footnotesize]
									\node[gauge,label=below:{$1$},fill=blue] (uN3) {};
									\node[gauge,label=below:{$N-3$}] (uN) [left=24mm of uN3]{};
									\node (dotsB) [left=6mm of uN] {$\cdots$};
									\node[gauge,label=below:{$2$}] (u2B) [left=6mm of dotsB] {};
									\node[gauge,label=below:{$1$}] (u1B) [left=6mm of u2B] {};
									\draw[thick] (u1B) -- (u2B) -- (dotsB) -- (uN);
									\draw[thick] (uN) -- node[above,sloped,midway] {$N-2$} (uN3);
									\node (U1d) [right=30mm of uN3] {$/\U(1)$};
									\node (Hfree) [below=8mm of dotsB] {$+\frac{N-3}{2}$ free hypermultiplets,};
								\end{tikzpicture}
							\end{equation}
							which is the expected 3d mirror of $D_N(\SU(N-2))$. Finally, we can deform this quiver by eliminating the remaining full tail. Each step increases the number of free hypermultiplets by one and decreases the multiplicity of the edge connecting the tail to the single $\U(1)$ by one. The resulting 3d mirror consists solely of the following free hypermultiplets:
							\begin{equation}
								H_\text{\tiny free} = \frac{N-3}{2}+\sum_{i=1}^{N-3} i = \frac{1}{2}(N-3)(N-1)\coma
							\end{equation}
							which matches the number of free hypermultiplets for the 3d mirror of $(A_{N-1},A_{N-3})$ for odd $N$ \cite[(6.35)-(6.36)]{Carta:2021whq}.
							
							\subsection{Summary of the Deformation Procedure}
							
							The examples in this and the preceding sections motivate the general strategy for deforming a parent quiver into the mirror of a general $D_p(\SU(N))$ theory. Once Tail B and Tail A, as defined in \eqref{def:TailATailB}, are obtained by deforming the parent tails corresponding to the partitions
							\begin{equation}
								\left[1^{N-x}\right]\coma \quad \left[(p-1)^x,N-px\right]\coma
							\end{equation}
							as explained in Section~\ref{sec:AnsatzforDpSUN}, the next step is to merge the abelian nodes in the bouquet until their number equals $\mu = \GCD(p,N)$.\footnote{This step is necessary only when $\mu=1$; otherwise, the number of $\U(1)$s is already equal to $\mu$.} Finally, the remaining full tails are eliminated one by one via deformations involving the $\mu$ abelian nodes of the bouquet. These deformations are responsible for increasing the multiplicity of the edges connecting the bouquet to Tails A and B, as well as the edges within the bouquet itself. If, during this process, the abelian nodes become connected to the tails via multiple edges, removing those tails will generate free hypermultiplets, as demonstrated in this section. Although we do not provide a formal proof, the examples presented offer substantial evidence that this procedure correctly reproduces the 3d mirror of the corresponding $D_p(\SU(N))$ theory.

							\acknowledgments 
							
							The authors thank Hamza Ahmed, Rafael \'Alvarez-Garc\'ia, Guillermo Arias-Tamargo, Florent Baume, Guido Bonori, Markus Dierigl, Monica Jinwoo Kang, Craig Lawrie, Lorenzo Mansi and Paul-Konstantin Oehlmann for discussions. W. H. also acknowledges the DESY Theory Group, Hamburg, in particular Craig Lawrie, and the Abdus Salam Centre for Theoretical Physics, Imperial College London, especially Amihay Hanany, for hospitality during the completion of this project. The work of S. G. is supported by the INFN grant ``Per attività di formazione per sostenere progetti di ricerca'' (GRANT 73/STRONGQFT). W. H. and
							N. M. are partially supported by the MUR-PRIN grant No. 2022NY2MXY (Finanziato dall’Unione europea -- Next Generation EU, Missione 4 Componente 1 CUP H53D23001080006, I53D23001330006). A. M. is supported in part by the DOE grant DE-SC0017647.
							
							\appendix
							
							\section{Review of \texorpdfstring{$D_p(\SU(N))$}{DpSUN} and Their 3d Mirror Theories}
							\label{sec:DpSUmirror-review}
							
							In the following sections, we are going to review the constructions of the 3d mirror theories for $D_p(\SU(N))$ theories, which represent the end points of the FI deformations acting on certain star-shaped quivers.

							\subsection{3d Mirror Theories for \texorpdfstring{$p\geq N$}{p>=N}}
							\label{sec:DpSUNmirrorsp>=N}
							
							Let us consider the 3d mirror theories for $D_p(\SU(N))$ when $p\geq N$. These have been studied in \cite{Giacomelli:2020ryy} and can be constructed as follows:
							\begin{enumerate}
								\item We construct a complete graph with $\mu$ vertices such that each vertex is a $\U(1)$ gauge group, and they are connected by an edge with multiplicity
								\begin{equation}
									m_G= \fn(\fp - \fn) = \frac{1}{\mu^2}N(p-N)\fstop 
								\end{equation}
								\item We also add a tail 
								\begin{equation}\label{eq:N-1tail}
									\begin{tikzpicture}[baseline=0,font=\footnotesize]
										\node[gauge,label=below:{$1$}] (u1) {};
										\node[gauge,label=below:{$2$}] (u2) [right=6mm of u1] {};
										\node (dots) [right=6mm of u2] {$\cdots$};
										\node[gauge,label=below:{$N-1$}] (uN1) [right=6mm of dots]{};
										\draw[thick] (u1) -- (u2) -- (dots) -- (uN1);
									\end{tikzpicture}
								\end{equation}
								\item We connect the $\U(N-1)$ node of the tail to all the $\U(1)$s in the complete graph with an edge with multiplicity $\fn = \frac{N}{\mu}$. 
								\item The 3d mirror has also a number of free hypermultiplets given by
								\begin{equation}
									H_\text{\tiny free} = \frac{1}{2}\mu(\fn -1)(\fp-\fn-1)=\frac{1}{2\mu}(N-\mu)(p-N-\mu)\fstop
								\end{equation}
							\end{enumerate}
							We can schematically draw the 3d mirror quiver as
							\begin{equation}\label{eq:generalDpSU-p>=N}
								\begin{tikzpicture}[baseline=0,font=\footnotesize]
									\node (p0) at (0,0) {};
									\node[gauge,label=above:{$1$}] (p2) at (45:2cm) {};
									\node[gauge,label=below:{$1$}] (p5) at (315:2cm) {};
									\node (dotsp2) [right=1cm of p0] {$\vdots$};
									\node[gauge,label={[xshift=-0.2cm,yshift=-0.76cm]{$N-1$}}] (uN1) [left=1cm of p0] {};
									\node (dots) [left=6mm of uN1] {$\cdots$};
									\node[gauge,label=below:{$2$}] (u2) [left=6mm of dots] {};
									\node[gauge,label=below:{$1$}] (u1) [left=6mm of u2]{};
									\node (U1d) [right=25mm of p5] {$/\U(1)$};
									\draw[thick] (u1) -- (u2) -- (dots) -- (uN1);
									\draw[thick] (p2) to[bend left=30] node[midway,above,sloped] {$m_G$} (p5);
									\draw[thick] (uN1) -- node[midway,above] {$\fn$} (p2);
									\draw[thick] (uN1) -- node[midway,below] {$\fn$} (p5);
									\draw[thick] [decorate,decoration={brace,amplitude=5pt},xshift=0cm, yshift=0cm]
									([xshift=10mm]p2.north) -- ([xshift=10mm]p5.south) node [black,pos=0.5,xshift=0.4cm] 
									{$\mu$};
									\node [below=15mm of u1, xshift=20mm] {$+H_\text{\tiny free}$ free hypermultiplets.};
								\end{tikzpicture}
							\end{equation}
							The mirror of $(A_{p-N-1},A_{N-1})$ is simply the complete graph of $\U(1)$s together with $H_{\text{\tiny free}}$ hypermultiplets \cite{Giacomelli:2020ryy}. 
							
							\subsection{3d Mirror Theories for \texorpdfstring{$p< N$}{p<N}}
							\label{sec:DpSUNmirrorsp<=N}
							
							In order to construct the mirror theories for $D_p(\SU(N))$ with $p\leq N$, we introduce, together with \eqref{eq:notationDpSU}, the following quantities:
							\begin{equation}
								x=\left\lfloor \frac{N}{p}\right\rfloor\aand M = N-(x+1)\fstop
								\label{eq:DpSUquantities}
							\end{equation}
							The mirror theory is given by:
							\begin{enumerate}
								\item A complete graph with $\mu$ $\U(1)$ nodes each with multiplicity
								\begin{equation} \label{defmAmBmG}
									m_{G} = \underbrace{\left(\fp (1+x)-\fn\right)}_{m_A}\underbrace{(\fn-\fp x)}_{m_B}\fstop
								\end{equation}
								\item We construct two tails
								\begin{equation} \label{def:TailATailB}
									\begin{tikzpicture}[baseline=-25,font=\footnotesize]
										\node[gauge,label=below:{$p-1$}] (u1) {};
										\node[gauge,label=below:{$2(p-1)$}] (u2) [right=12mm of u1] {};
										\node (dots) [right=6mm of u2] {$\cdots$};
										\node[gauge,label=below:{$(x-1)(p-1)$}] (uN1) [right=6mm of dots]{};
										\node[gauge,label=below:{$x(p-1)$}] (uN2) [right=18mm of uN1]{};
										\draw[thick] (u1) -- (u2) -- (dots) -- (uN1) -- (uN2);
										\node [left=6mm of u1] {Tail A:};
										\node[gauge,label=below:{$1$}] (u1B) [below=12mm of u1] {};
										\node[gauge,label=below:{$2$}] (u2B) [right=12mm of u1B] {};
										\node (dotsB) [right=6mm of u2B] {$\cdots$};
										\node[gauge,label=below:{$M-1$}] (uN1B) [right=6mm of dotsB]{};
										\node[gauge,label=below:{$M$}] (uN2B) [right=18mm of uN1B]{};
										\draw[thick] (u1B) -- (u2B) -- (dotsB) -- (uN1B) -- (uN2B);
										\node [left=6mm of u1B] {Tail B:};
									\end{tikzpicture}
								\end{equation}
								\item Each $\U(1)$ node of the complete graph is attached to the $\U(x(p-1))$ node of Tail A with an edge with multiplicity $m_A$.
								\item Each $\U(1)$ node of the complete graph is attached to the $\U(M)$ node of Tail B with an edge with multiplicity $m_B$.
								\item The node $\U(x(p-1))$ node of Tail A is attached to the $\U(M)$ node of Tail B with an edge with multiplicity one.
								\item There are 
								\begin{equation}\label{eq:freehypers-p<=N}
									\begin{split}
										H_\text{\tiny free} &= \frac{1}{2}\mu (\fn -x\fp -1)(\fp(1+x)-\fn -1) = \frac{1}{2\mu}\left(N-p x-\mu\right)\left(p(1+x)-N-\mu \right)\\
										&=\frac{1}{2}\mu (m_B -1)(m_A -1) = \frac{1}{2}\left(\mu(m_G+1)-p\right)
									\end{split}
								\end{equation}
								free hypermultiplets. 
							\end{enumerate}
							
							Once again, we can draw the quiver schematically as follows: 
							\begin{equation}\label{eq:generalDpSU-p<=N}
								\begin{tikzpicture}[baseline=30,font=\footnotesize]
									\node[gauge,label=below:{$1$}] (uB1) {};
									\node[gauge,label=below:{$2$}] (uB2) [right=6mm of uB1] {};
									\node (dotsB1) [right=6mm of uB2] {$\cdots$};
									\node[gauge,label=below:{$M$}] (uBM) [right=6mm of dotsB1] {};
									\node[gauge,label=below:{$x(p-1)$}] (uN1) [right=24mm of uBM]{};
									\node[gauge,label=below:{$(x-1)(p-1)$}] (uN2) [right=18mm of uN1]{};
									\node (dotsA1) [right=10mm of uN2] {$\cdots$};
									\node[gauge,label=below:{$p-1$}] (uNp) [right=10mm of dotsA1]{};
									\node (U1d) [right=10mm of uNp] {$/\U(1)$};
									\node[gauge,label=above:{$1$}] (u1) [above=18mm of uBM] {};
									\node[gauge,label=above:{$1$}] (u12) [above=18mm of uN1] {};
									\node [below= 8mm of uB2,xshift=12mm] {$+H_\text{\tiny free}$ free hypermultiplets.};
									\node (dotsu1) [right=7mm of u1] {$\cdots$};
									\draw[thick] [decorate,decoration={brace,amplitude=5pt},xshift=0cm, yshift=0cm]
									([yshift=8mm]u1.west) -- ([yshift=8mm]u12.east) node [black,pos=0.5,yshift=0.4cm] 
									{$\mu$};
									\draw[thick] (uB1) -- (uB2) -- (dotsB1) -- (uBM) -- (uN1) -- (uN2) -- (dotsA1) -- (uNp);
									\draw[thick] (u1) to[bend left=20] node[midway,above,sloped] {$m_G$} (u12);
									\draw[thick] (uBM) -- node[midway,above,sloped] {$m_B$} (u1);
									\draw[thick] (uBM) -- node[pos=0.3,above, sloped] {$m_B$} (u12);
									\draw[thick] (uN1) -- node[pos=0.3,above, sloped] {$m_A$} (u1);
									\draw[thick] (u12) -- node[midway,above,sloped] {$m_A$} (uN1);
								\end{tikzpicture}
							\end{equation}
							
							\section{From Orbi-instantons to Generic Star-shaped Quivers}
							\label{sec:OItoStarShaped}
							
							In \cite{Giacomelli:2024ycb}, it was shown that combinations of the Fayet-Iliopoulos (FI) deformations reviewed in Section~\ref{sec:FIDeformations-review} allow for the construction of any Type A class $\mathcal{S}$ theory with regular punctures from an orbi-instanton theory. While we refer the reader to \cite{Giacomelli:2024ycb} for a comprehensive study of the deformations that map an orbi-instanton theory to a generic star-shaped quiver, this Appendix will briefly review the procedure.
							
							\subsection{Deformations from Orbi-instanton Theories to Star-shaped Quivers}
							
							Let us consider a generic magnetic quiver for an orbi-instanton theory of the form \cite{Mekareeya:2017jgc} (see also \cite{Cabrera:2019izd}):
							\begin{equation}\label{eq:generalE8quiv}
								\begin{tikzpicture}[baseline=7,font=\footnotesize]
									\node (dots) {$\cdots$};
									\node[gauge, label=below:{$k$}] (k) [right=6mm of dots] {};
									\node[gauge, label=below:{$A$}] (A) [right=6mm of k] {};
									\node[gauge, label=below:{$B$}] (B) [right=6mm of A] {};
									\node[gauge, label=below:{$C$}] (C) [right=6mm of B] {};
									\node[gauge, label=below:{$D$}] (D) [right=6mm of C] {};
									\node[gauge, label=below:{$E$}] (E) [right=6mm of D] {};
									\node[gauge, label=below:{$F$}] (F) [right=6mm of E] {};
									\node[gauge, label=below:{$G$}] (G) [right=6mm of F] {};
									\node[gauge, label=below:{$H$}] (H) [right=6mm of G] {};
									\node[gauge, label=above:{$L$}] (L) [above=4mm of F] {};
									\draw[thick] (dots.east) -- (k.west);
									\draw[thick] (k.east) -- (A.west);
									\draw[thick] (A.east) -- (B.west);
									\draw[thick] (B.east) -- (C.west);
									\draw[thick] (C.east) -- (D.west);
									\draw[thick] (D.east) -- (E.west);
									\draw[thick] (E.east) -- (F.west);
									\draw[thick] (F.east) -- (G.west);
									\draw[thick] (G.east) -- (H.west);
									\draw[thick] (F.north) -- (L.south);
								\end{tikzpicture}
							\end{equation}
							We assume that the $\U(F)$ node is balanced, i.e.,
							\begin{equation}\label{eq:n6rankconstraint}
								2F = E+G+L\fstop
							\end{equation}
							We can now perform FI deformations involving node $F$ and its surrounding nodes $E$, $G$, and $L$, whose ranks are constrained by \eqref{eq:n6rankconstraint}. Implementing the procedure discussed in \cite{Giacomelli:2024ycb} and reviewed in Section \ref{sec:FIDeformations-review} leads to the quiver
							\begin{equation}\label{eq:E7quiverFromn6=0}
								\begin{tikzpicture}[baseline=7,font=\footnotesize]
									\node (dots) {$\cdots$};
									\node[gauge, label=below:{$k$}] (k) [right=6mm of dots] {};
									\node[gauge, label=below:{$A$}] (A) [right=6mm of k] {};
									\node[gauge, label=below:{$B$}] (B) [right=6mm of A] {};
									\node[gauge, label=below:{$C$}] (C) [right=6mm of B] {};
									\node[gauge, label=below:{$D$}] (D) [right=6mm of C] {};
									\node[gauge, label=above:{$G+E-F$}] (E) [right=9mm of D] {};
									\node[gauge, label=below:{$H$}] (F) [right=9mm of E] {};
									\node[gauge, label=above:{$F-E$}] (G) [right=6mm of F] {};
									\node[gauge, label=above:{$F-G$}] (FG) [above=4mm of D] {};
									\draw[thick] (dots.east) -- (k.west);
									\draw[thick] (k.east) -- (A.west);
									\draw[thick] (A.east) -- (B.west);
									\draw[thick] (B.east) -- (C.west);
									\draw[thick] (C.east) -- (D.west);
									\draw[thick] (D.east) -- (E.west);
									\draw[thick] (E.east) -- (F.west);
									\draw[thick] (F.east) -- (G.west);
									\draw[thick] (D.north) -- (FG.south);
								\end{tikzpicture}
							\end{equation}
							This is an $E_7$-shaped quiver with a single tail, which is inherited from the parent quiver. This resulting quiver can be further deformed into an $E_6$-shaped quiver. For example, assuming $B \geq F-G$, we can activate FI parameters for these nodes. If there exists a node in the tail with rank $\tilde{k}=B-F+G$, the FI parameter must also be activated for that node. The resulting quiver after this deformation is
							\begin{equation}
								\begin{tikzpicture}[baseline=7,font=\footnotesize]
									\node (dots) {$\cdots$};
									\node[gauge, label=below:{$k-\tilde{k}$}] (k) [right=3mm of dots] {};
									\node[gauge,label=above:{$A-\tilde{k}$}] (A) [right=9mm of k] {};
									\node[gauge,label=below:{$F-G$}] (B) [right=9mm of A] {};
									\node[gauge,label=below:{$E-F+G$}] (C) [right=16mm of B] {};
									\node[gauge,label=above:{$D-F+G$}] (D) [right=9mm of C] {};
									\node[gauge,label=below:{$C-F+G$}] (E) [right=9mm of D] {};
									\node[gauge,label=left:{$H$}] (F) [above=4mm of C] {};
									\node[gauge,label=above:{$F-E$}] (G) [above=4mm of F] {};
									\node[gauge, label=below:{$\tilde{k}$}] (tk) [right=12mm of E] {};
									\node (dots2) [right=3mm of tk] {$\cdots$};
									\draw[thick] (dots.east) -- (k.west);
									\draw[thick] (k.east) -- (A.west);
									\draw[thick] (A.east) -- (B.west);
									\draw[thick] (B.east) -- (C.west);
									\draw[thick] (C.east) -- (D.west);
									\draw[thick] (D.east) -- (E.west);
									\draw[thick] (C.north) -- (F.south);
									\draw[thick] (F.north) -- (G.south);
									\draw[thick] (E.east) -- (tk.west);
									\draw[thick] (tk.east) -- (dots2.west);
								\end{tikzpicture}
								\label{eq:BHdef-E71tailtoE62tail}
							\end{equation}
							This process generates an $E_6$-shaped quiver with two tails. One tail originates from the original tail of the $E_7$-shaped quiver, with all its ranks reduced by $\tilde{k}$, while the second tail is new and terminates at a node of rank $\tilde{k}$. Different types of deformations can produce an $E_6$-shaped quiver with three tails, of the general form
							\begin{equation}\label{eq:genE6quiv-2}
								\begin{tikzpicture}[baseline=7,font=\footnotesize]
									\node (T1) {$T_{\rho_1,\rho_2,\ldots,\rho_k}$};
									\node[gauge,label=below:{$kN$}] (Nk) [right=6mm of T1] {};
									\node (T2) [right=6mm of Nk]{$T_{\rho_1',\rho_2',\ldots,\rho_{k'}'}$};
									\node[gauge,label=left:{$N$}] (N) [above=4mm of Nk] {};
									\node (T3) [above=4mm of N] {$T_\rho$};
									\draw[thick] (T1.east) -- (Nk.west);
									\draw[thick] (Nk.east) -- (T2.west);
									\draw[thick] (Nk.north) -- (N.south);
									\draw[thick] (N.north) -- (T3.south);
								\end{tikzpicture}
							\end{equation}
							Here, following the notation in \cite{Giacomelli:2024ycb}, we define $T_{\rho_1,\rho_2,\ldots,\rho_k}$ as a tail
							\begin{equation}\label{eq:Trho123}
								\begin{tikzpicture}[baseline=0,font=\footnotesize]
									\node (T1) {$T_{\rho_1}$};
									\node[gauge,label=below:{$N$}] (N) [right=6mm of T1] {};
									\node (Y2) [right=6mm of N] {$Y_2$};
									\node[gauge,label=below:{$2N$}]  (N2) [right=6mm of Y2] {};
									\node (Y3) [right=6mm of N2] {$Y_3$};
									\node (dots) [right=6mm of Y3] {$\cdots$};
									\node[gauge,label=below:{$(k-1)N$}]  (Nk) [right=6mm of dots] {};
									\node (Yk) [right=6mm of Nk] {$Y_k$};
									\draw[thick] (T1) -- (N) -- (Y2) -- (N2) -- (Y3) -- (dots) -- (Nk) -- (Yk);
								\end{tikzpicture}
							\end{equation}
							where $Y_i$ are $T_{\rho_i}$ tails with their node ranks shifted by $(i-1)N$. For example, if we consider $\rho_i = \left[1^N\right]$, the quiver for $T_{\rho_i}$ is
							\begin{equation}
								\begin{tikzpicture}[baseline=0,font=\footnotesize]
									\node[gauge,label=below:{$1$}] (u1) {};
									\node[gauge,label=below:{$2$}] (u2) [right=6mm of u1] {};
									\node (dots) [right=6mm of u2] {$\cdots$};
									\node[gauge,label=below:{$N-1$}] (uN1) [right=6mm of dots]{};
									\node[flavor,label=below:{$N$}] (fN) [right=9mm of uN1] {};
									\draw[thick] (u1) -- (u2) -- (dots) -- (uN1) -- (fN);
								\end{tikzpicture}
							\end{equation}
							The corresponding $Y_i$ is then
							\begin{equation}
								\begin{tikzpicture}[baseline=0,font=\footnotesize]
									\node[gauge,label=below:{$1+(i-1)N$}] (u1) {};
									\node[gauge,label=below:{\qquad $2+(i-1)N$}] (u2) [right=18mm of u1] {};
									\node (dots) [right=15mm of u2]{$\cdots$};
									\node[gauge,label=below:{$iN-1$}] (uM1) [right=9mm of dots]{};
									\node[flavor,label=below:{$iN$}] (fM) [right=9mm of uM1] {};
									\draw[thick] (u1) -- (u2) -- (dots) -- (uM1) -- (fM);
								\end{tikzpicture}
							\end{equation}
							and its terminal flavor node is gauged with the first node of $Y_{i+1}$. If we sequentially activate FI parameters in \eqref{eq:genE6quiv-2} at pairs of nodes with ranks $N, 2N, \ldots$, we obtain a star-shaped quiver of the form
							\begin{equation} \label{genericssq}
								\begin{tikzpicture}[baseline=0,font=\footnotesize]
									\node[gauge,label=left:{$N$}] (N) {};
									\node (T1) [above right=9mm of N] {$T_{\rho_k}$};
									\node (dots1) [above=7mm of N] {$\cdots$};
									\node (Tk) [above left=9mm of N] {$T_{\rho_1}$};
									\node (T1p) [below left=9mm of N] {$T_{\rho_1'}$};
									\node (dots2) [below=7mm of N] {$\cdots$};
									\node (Tkp) [below right=9mm of N] {$T_{\rho_{k'}'}$};
									\node (T3) [right=9mm of N] {$T_{\rho}$};
									\draw[thick] (N) -- (T1);
									\draw[thick] (N) -- (Tk);
									\draw[thick] (N) -- (T1p);
									\draw[thick] (N) -- (Tkp);
									\draw[thick] (N) -- (T3);
									\draw[thick] [decorate,decoration={brace,amplitude=5pt},xshift=0cm, yshift=0cm]
									([yshift=5mm]Tk.west) -- ([yshift=5mm]T1.east) node [black,pos=0.5,yshift=0.4cm] 
									{$k$ tails};
									\draw[thick] [decorate,decoration={mirror,brace,amplitude=5pt},xshift=0cm, yshift=0cm]
									([yshift=-5mm]T1p.west) -- ([yshift=-5mm]Tkp.east) node [black,pos=0.5,yshift=-0.5cm] 
									{$k'$ tails};
								\end{tikzpicture}
							\end{equation}
							This procedure yields a generic star-shaped quiver with an arbitrary number of tails. 
							
							\subsection{Inverse Algorithm from Star-shaped Quivers to Orbi-instanton Theories}
							
							The pattern of FI deformations that leads from an orbi-instanton theory to a generic star-shaped quiver can be inverted. This allows one to predict the parent $E_6$-shaped quiver for a given star-shaped quiver. The inverse transformation generically produces a quiver with underbalanced nodes, which can be made balanced by applying Seiberg-like dualities \cite{Yaakov:2013fza} on the underbalanced node where the FI parameter is turned on \cite{Assel:2017jgo, Giacomelli:2023zkk}. The strategy for a generic star-shaped quiver, such as \eqref{genericssq}, is to propose a candidate $E_6$-shaped parent theory of the form \eqref{eq:genE6quiv-2} and then apply dualities to the underbalanced nodes until a ``good" quiver is obtained. The result is the parent $E_6$-shaped quiver from which the star-shaped quiver in question originates. This procedure can be iterated to trace the lineage back to the parent orbi-instanton mirror quiver, applying dualities at each step of the reconstruction. A full discussion with further details can be found in \cite{Giacomelli:2024ycb}. 
							
							Since this inverse algorithm is deterministic (up to dualities), the main focus of this work has been to find a prescription for the parent star-shaped quiver from which a $D_p(\SU(N))$ theory can be obtained via FI deformations. This star-shaped quiver can then be traced back to the corresponding orbi-instanton theory by implementing the algorithm described in \cite{Giacomelli:2024ycb}.

							\bibliographystyle{JHEP}
							\bibliography{mybib}

						\end{document}